\documentclass[prc,twocolumn,superscriptaddress,showpacs,amssymb,amsmath,amsfonts,aps]{revtex4}
\usepackage{graphicx}
\usepackage{dcolumn}
\usepackage{bm}
\usepackage{color}

\begin{document}



\newcommand*{\ASU }{ Arizona State University, Tempe, Arizona 85287-1504} 
\affiliation{\ASU } 

\newcommand*{\SACLAY }{ CEA-Saclay, Service de Physique Nucl\'eaire, F91191 Gif-sur-Yvette, France} 
\affiliation{\SACLAY } 

\newcommand*{\UCLA }{ University of California at Los Angeles, Los Angeles, California  90095} 
\affiliation{\UCLA } 

\newcommand*{\CMU }{ Carnegie Mellon University, Pittsburgh, Pennsylvania 15213} 
\affiliation{\CMU } 

\newcommand*{\CUA }{ Catholic University of America, Washington, D.C. 20064} 
\affiliation{\CUA } 

\newcommand*{\CNU }{ Christopher Newport University, Newport News, Virginia 23606} 
\affiliation{\CNU } 

\newcommand*{\UCONN }{ University of Connecticut, Storrs, Connecticut 06269} 
\affiliation{\UCONN } 


\newcommand*{\ECOSSEE }{ Edinburgh University, Edinburgh EH9 3JZ, United Kingdom} 
\affiliation{\ECOSSEE } 

\newcommand*{\FIU }{ Florida International University, Miami, Florida 33199} 
\affiliation{\FIU } 

\newcommand*{\FSU }{ Florida State University, Tallahassee, Florida 32306} 
\affiliation{\FSU } 


\newcommand*{\GWU }{ The George Washington University, Washington, DC 20052} 
\affiliation{\GWU } 

\newcommand*{\ECOSSEG }{ University of Glasgow, Glasgow G12 8QQ, United Kingdom} 
\affiliation{\ECOSSEG } 

\newcommand*{\IDAHO }{ Idaho State University, Pocatello, Idaho, 83209} 
\affiliation{\IDAHO }

\newcommand*{\INFNFR }{ INFN, Laboratori Nazionali di Frascati, Frascati, Italy} 
\affiliation{\INFNFR } 

\newcommand*{\INFNGE }{ INFN, Sezione di Genova, 16146 Genova, Italy} 
\affiliation{\INFNGE } 

\newcommand*{\ORSAY }{ Institut de Physique Nucl\'eaire ORSAY, F91406 Orsay, France} 
\affiliation{\ORSAY } 

\newcommand*{\ITEP }{ Institute of Theoretical and Experimental Physics, Moscow, 117259, Russia} 
\affiliation{\ITEP } 

\newcommand*{\JMU }{ James Madison University, Harrisonburg, Virginia 22807} 
\affiliation{\JMU } 

\newcommand*{\KYUNGPOOK }{ Kyungpook National University, Daegu 702-701, South Korea} 
\affiliation{\KYUNGPOOK } 

\newcommand*{\MIT }{ Massachusetts Institute of Technology, Cambridge, Massachusetts  02139} 
\affiliation{\MIT } 

\newcommand*{\UMASS }{ University of Massachusetts, Amherst, Massachusetts  01003} 
\affiliation{\UMASS } 

\newcommand*{\UNH }{ University of New Hampshire, Durham, New Hampshire 03824} 
\affiliation{\UNH } 

\newcommand*{\NSU }{ Norfolk State University, Norfolk, Virginia 23504} 
\affiliation{\NSU } 

\newcommand*{\OHIOU }{ Ohio University, Athens, Ohio  45701} 
\affiliation{\OHIOU } 

\newcommand*{\ODU }{ Old Dominion University, Norfolk, Virginia 23529} 
\affiliation{\ODU } 

\newcommand*{\PENN }{ Penn State University, University Park, Pennsylvania 16802}
\affiliation{\PENN } 

\newcommand*{\PITT }{ University of Pittsburgh, Pittsburgh, Pennsylvania 15260}
\affiliation{\PITT } 

\newcommand*{\RPI }{ Rensselaer Polytechnic Institute, Troy, New York 12180-3590} 
\affiliation{\RPI } 

\newcommand*{\RICE }{ Rice University, Houston, Texas 77005-1892} 
\affiliation{\RICE } 

\newcommand*{\URICH }{ University of Richmond, Richmond, Virginia 23173} 
\affiliation{\URICH } 


\newcommand*{\SCAROLINA }{ University of South Carolina, Columbia, South Carolina 29208} 
\affiliation{\SCAROLINA } 

\newcommand*{\UTEP }{ University of Texas at El Paso, El Paso, Texas 79968} 
\affiliation{\UTEP } 

\newcommand*{\JLAB }{ Thomas Jefferson National Accelerator Facility, Newport News, Virginia 23606} 
\affiliation{\JLAB } 

\newcommand*{\UNIONC }{ Union College, Schenectady, NY 12308} 
\affiliation{\UNIONC } 

\newcommand*{\VT }{ Virginia Polytechnic Institute and State University, Blacksburg, Virginia   24061} 
\affiliation{\VT } 

\newcommand*{\VIRGINIA }{ University of Virginia, Charlottesville, Virginia 22901} 
\affiliation{\VIRGINIA } 

\newcommand*{\WM }{ College of William and Mary, Williamsburg, Virginia 23187} 
\affiliation{\WM } 

\newcommand*{\YEREVAN }{ Yerevan Physics Institute, 375036 Yerevan, Armenia} 
\affiliation{\YEREVAN } 

\newcommand*{\deceased }{ Deceased} 


\newcommand*{\NOWNCATU }{ North Carolina Agricultural and Technical State University, Greensboro, NC 27411}

\newcommand*{\NOWECOSSEG }{ University of Glasgow, Glasgow G12 8QQ, United Kingdom}

\newcommand*{\NOWSACLAY }{ CEA-Saclay, Service de Physique Nucl\'eaire, F91191 Gif-sur-Yvette, France}

\newcommand*{\NOWJLAB }{ Thomas Jefferson National Accelerator Facility, Newport News, Virginia 23606}

\newcommand*{\NOWINFNFR }{ INFN, Laboratori Nazionali di Frascati, Frascati, Italy}

\newcommand*{\NOWSCAROLINA }{ University of South Carolina, Columbia, South Carolina 29208}

\newcommand*{\NOWFIU }{ Florida International University, Miami, Florida 33199}

\newcommand*{\NOWINDSTRA }{ Systems Planning and Analysis, Alexandria, Virginia 22311}

\newcommand*{\NOWASU }{ Arizona State University, Tempe, Arizona 85287-1504}

\newcommand*{\NOWCISCO }{ Cisco, Washington, DC 20052}

\newcommand*{\NOWUK }{ University of Kentucky, LEXINGTON, KENTUCKY 40506}

\newcommand*{\NOWWM }{ College of William and Mary, Williamsburg, Virginia 23187-8795}

\newcommand*{\NOWRPI }{ Rensselaer Polytechnic Institute, Troy, New York 12180-3590}

\newcommand*{\NOWROMA }{ Universita' di ROMA III, 00146 Roma, Italy} 

\newcommand*{\NOWUNCW }{ North Carolina}

\newcommand*{\NOWHAMPTON }{ Hampton University, Hampton, VA 23668}

\newcommand*{\NOWMOSCOW }{ Moscow State University, General Nuclear Physics Institute, 119899 Moscow, Russia}

\newcommand*{\NOWTulane }{ Tulane University, New Orleans, Lousiana  70118}

\newcommand*{\NOWOHIOU }{ Ohio University, Athens, Ohio  45701}

\newcommand*{\NOWKYUNGPOOK }{ Kungpook National University, Taegu 702-701, South Korea}

\newcommand*{\NOWCUA }{ Catholic University of America, Washington, D.C. 20064}

\newcommand*{\NOWGEORGETOWN }{ Georgetown University, Washington, DC 20057}

\newcommand*{\NOWJMU }{ James Madison University, Harrisonburg, Virginia 22807}

\newcommand*{\NOWURICH }{ University of Richmond, Richmond, Virginia 23173}

\newcommand*{\NOWCALTECH }{ California Institute of Technology, Pasadena, California 91125}

\newcommand*{\NOWVIRGINIA }{ University of Virginia, Charlottesville, Virginia 22901}

\newcommand*{\NOWYEREVAN }{ Yerevan Physics Institute, 375036 Yerevan, Armenia}

\newcommand*{\NOWUMASS }{ University of Massachusetts, Amherst, Massachusetts  01003}

\newcommand*{\NOWRICE }{ Rice University, Houston, Texas 77005-1892}

\newcommand*{\NOWINFNGE }{ INFN, Sezione di Genova, 16146 Genova, Italy}

\newcommand*{\NOW }{ }

\newcommand*{\NOWBATES }{ MIT-Bates Linear Accelerator Center, Middleton, MA 01949}

\newcommand*{\NOWODU }{ Old Dominion University, Norfolk, Virginia 23529}

\newcommand*{\NOWFSU }{ Florida State University, Tallahassee, Florida 32306}

\newcommand*{\NOWGEISSEN }{ Physikalisches Institut der Universit\"at Giessen, 35392 Giessen, Germany} 

\newcommand*{\NOWVSU }{ Virginia State University, Petersburg,Virginia 23806}

\newcommand*{\NOWORST }{ Oregon State University, Corvallis, Oregon 97331-6507}

\newcommand*{\NOWCMU }{ Carnegie Mellon University, Pittsburgh, Pennsylvania 15213}

\newcommand*{\NOWMIT }{ Massachusetts Institute of Technology, Cambridge, Massachusetts  02139-4307}

\newcommand*{\NOWGWU }{ The George Washington University, Washington, DC 20052}

\title{Complete measurement of three-body photodisintegration of $^3$He for photon energies between 0.35 and 1.55 GeV}
\author{S.~Niccolai}
     \altaffiliation[Current address:]{\ORSAY}
     \affiliation{\SACLAY}
     \affiliation{\GWU}
\author{G.~Audit}
     \affiliation{\SACLAY}
\author{B.L.~Berman}
     \affiliation{\GWU}
\author{J.M.~Laget}
     \affiliation{\SACLAY}
\author{S.~Strauch}
     \affiliation{\GWU}
\author{G.~Adams}
     \affiliation{\RPI}
\author{A.~Afanasev}
     \affiliation{\YEREVAN}
\author{P.~Ambrozewicz}
     \affiliation{\FIU}
\author{M.~Anghinolfi}
     \affiliation{\INFNGE}
\author{J.R.M.~Annand}
     \affiliation{\ECOSSEG}
% ...

\collaboration{The CLAS Collaboration}
     \noaffiliation




\date{\today}

\begin{abstract}
The three-body photodisintegration of $^3{\mbox{He}}$ has been
measured with the CLAS detector at Jefferson Lab, using tagged photons
of energies between 0.35 GeV and 1.55 GeV. The large acceptance of the
spectrometer allowed us for the first time to cover a wide momentum
and angular range for the two outgoing protons. Three kinematic
regions dominated by either two- or three-body contributions have been
distinguished and analyzed. The measured cross sections have been
compared with results of a theoretical model, which, in
certain kinematic ranges, have been found to be in reasonable
agreement with the data.
\end{abstract}


\maketitle

\section{\label{sec:intro}INTRODUCTION}

The study of the electromagnetic properties of the $^3$He nucleus is
the optimal starting point to assess the importance of many-body
interactions between nucleons in nuclei \cite{barry_book,fb17}.
In particular, in the $\gamma^3{\mbox{He}} \rightarrow ppn$
reaction, three-nucleon currents dominate in certain regions of phase
space \cite{laget_d,laget1,laget1b}. In fact, a $pp$ pair has no
dipole moment with which to couple and the charge-exchange current
vanishes within a $pp$ pair, so that the one- and two-nucleon currents
are suppressed in those regions.  The small number of nucleons
involved makes possible kinematically complete experiments, and exact
Faddeev ground-state wave functions, as well as exact wave functions
for the continuum three-body final state at low energies (below the
pion-production threshold), are available \cite{glockle96,glockle02}.

Although the calculations of the $^3$He ground-state wave function
have reached a high level of accuracy in reproducing the bound-state
properties \cite{glockle96,glockle02}, the calculation of the
continuum three-nucleon wave function is less developed at higher
energy; a full treatment of the three-body photodisintegration of
$^3$He has been possible only at energies $E_{\gamma}\leq 300$ MeV.
As the energy increases, the number of partial waves and open channels
becomes very large and, so far, no calculations that are both exact
and complete have been done in the GeV region. Not only would a
very large computational effort be required to do so, but also a
treatment of the absorptive part of the nucleon-nucleon interaction
(coupling to other open channels that is not taken into account in
potential-based calculations) should be implemented.  

A different approach has been taken by Laget \cite{laget_a, laget_b,
laget_c, laget_d, laget_e}, who has employed a diagrammatic model for
the evaluation of the contribution of one-, two-, and three-body
mechanisms in the cross section for the photodisintegration of
$^3$He. Rather than relying on a partial-wave expansion, this approach
relies on the evaluation of the dominant graphs whose amplitudes are
related to one- and two-body elementary amplitudes. The
parametrization of these elementary amplitudes incorporates absorptive
effects due to the coupling with other channels, which become more and
more important as the energy increases.  The comparison of these model
predictions with experimental data provides us with a good starting
point to understand the nature of three-body interactions in $^3$He
for photon energies in the GeV region.  

At stake is the link with three-body forces. In the $^3$He ground
state, three-body forces involve the exchange of virtual mesons
between nucleons and the creation of virtual baryonic resonances. The
incoming photon can couple to each of these charged particles. Below
the pion-photoproduction threshold, all the particles remain virtual
and the corresponding three-body meson-exchange currents (MEC)
contribute only weakly to the cross section. When the photon energy
increases above the various meson- or resonance-production thresholds,
these virtual particles can become real -- they can propagate on-shell
\cite{barry_book}. The corresponding sequential scattering amplitudes
are considerably enhanced and can dominate certain well defined parts
of the phase space.  Kinematically complete experiments allow one to
isolate each of the dominant sequential rescattering amplitudes.  They
analytically reduce to three-body MEC at lower energy, and put
constraints on the corresponding three-body current.

Several low-energy ($<$100 MeV) experiments have been performed since
the publication of the results of the first measurement of the
three-body photodisintegration of $^3$He in 1964
\cite{berman1}. But only a few have been performed at intermediate
photon energies up to 800 MeV, in limited kinematics
\cite{audit91,sarty92,audit93} as well as with large-acceptance
detectors \cite{maruyama95,kolb96,audit97}. They show good agreement
with Laget's predictions provided that the $3N$ mechanisms, based on
sequential pion exchanges and $\Delta$-resonance formation, are
included in the calculations. Since these mechanisms dominate well
defined parts of the phase space, a better understanding of the nature
of many-body interactions requires one to perform a high-statistics
full $4\pi$ investigation, probing the three-body breakup process for
all angular and energy correlations of the three outgoing
nucleons. Also, the extension to the high-energy ($E_{\gamma}\geq1$
GeV) region, where no experiment has been performed until now, can be
expected to open a window on other kinds of many-body processes.

This paper reports on a measurement of the three-body
photodisintegration of $^3$He performed in Hall B at Jefferson Lab
\cite{tesi}. Photon energies between 0.35 GeV and 1.55 GeV were used,
and wide angular and momentum ranges for the outgoing particles were
covered. These features, along with the high statistics collected,
allow us to select the most interesting two- and three-body processes,
to compare their relative importance, and to determine their variation
with photon energy.

The experimental setup is described briefly in Sec.~\ref{sec:exp}, the
salient points of the data analysis in Sec.~\ref{sec:analysis}, and
the model calculation in Sec.~\ref{sec:res_full}. Our results for
several kinematic regions are presented in detail and compared with
the model calculation in Sec.~\ref{results}, and summarized in
Sec.~\ref{summary}.

\section{\label{sec:exp}EXPERIMENTAL SETUP}
The experiment was performed at the Thomas Jefferson National Accelerator Facility, in Hall B, using the CEBAF Large Acceptance Spectrometer (CLAS) \cite{clas} and the bremsstrahlung photon tagger \cite{tagger}. 
The electron beam energy was 1.645~GeV, corresponding to two passes of the CEBAF accelerator; the current was 10~nA during regular production runs and 0.1~nA during tagging-efficiency calibration runs.
The photon beam was produced by the electron beam striking the radiator, a thin layer ($\sim5\times10^{-5}$ radiation length) of gold deposited on a thin carbon backing, which was placed 50~cm before the entrance of the tagger magnet. The electrons interacting in the radiator were deflected by the magnetic field of the tagging magnet, and those with energy between 20\% and 95\% of the incident electron beam energy were detected by two layers of scintillators (E-counters, measuring the energy of the electron, and T-counters, measuring its time \cite{tagger})  placed in the magnet focal plane. Thus, photons in the energy range from 0.35 to 1.55~GeV were tagged. Two collimators were placed in the beamline between the tagger and the $^3$He target, in order to eliminate the tails from the photon beam and to give a small and well defined beam spot on the target.
The data were obtained using a cylindrical cryogenic target, 18~cm long and 4~cm in diameter, filled with liquid $^3$He and positioned approximately 20~m downstream of the tagger radiator in the center of the CLAS. 
A lead-glass total absorption counter (TAC), almost 100\% efficient, placed approximately 20 m downstream from the center of the CLAS detector, measured the tagging efficiency during low-flux calibration runs. 

The CLAS is a magnetic toroidal spectrometer in which the magnetic
field is generated by six superconducting coils. The six azimuthal
sectors are individually equipped with drift chambers for track
reconstruction, scintillation counters for time-of-flight measurement,
\v{C}erenkov counters for electron-pion discrimination, and
electromagnetic calorimeters to identify electrons and neutrals. 
The polarization of the CLAS torus was set to bend 
the negatively charged particles toward the beam line.
In order to achieve a good compromise between momentum resolution and
negative-particle acceptance (required by other simultaneous
experiments) the magnetic field of the CLAS was set to slightly less
than half of its maximum value, corresponding to a torus current of
1920 A. A coincidence 
between the tagger and the time-of-flight scintillators defined 
the Level-1 trigger for accepting the hadronic
events. For the first time in CLAS, a Level-2 trigger, which selected
the events from among those passed through Level-1 that have at least
one ``likely track'' in the drift chambers, was also used
\cite{clas}.  More than a billion events of production data were
obtained with $^3$He (plus a few million events taken with the target
empty), at a data-acquisition rate slightly greater than 3 kHz.

\section{\label{sec:analysis}DATA REDUCTION AND ANALYSIS}

\subsection{\label{sec:ch_id}Channel Identification}
\begin{figure}[htbp]
\includegraphics[scale=0.45]{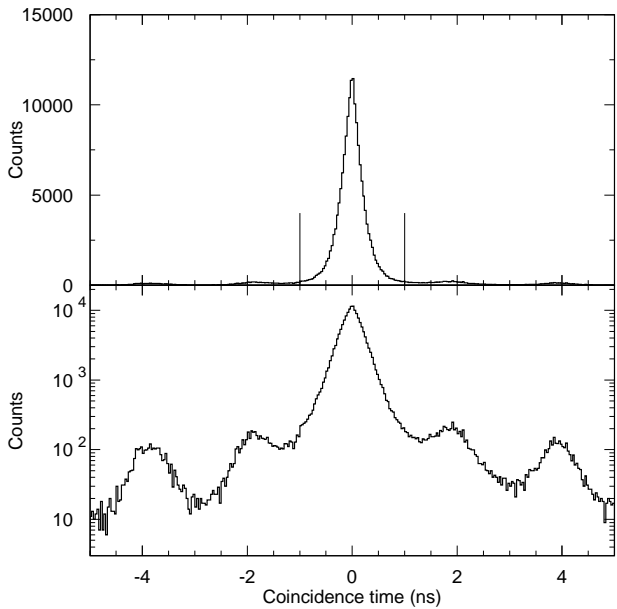}
\caption{\label{fig:coinc_time} Coincidence time for a subset of the
  raw data. The vertical lines in the upper panel indicate the time
  window for accepted events. Random coincidences from neighboring
  beam pulses are visible in the lower panel.}
\end{figure}

\begin{figure}[htbp]
\includegraphics[scale=0.45]{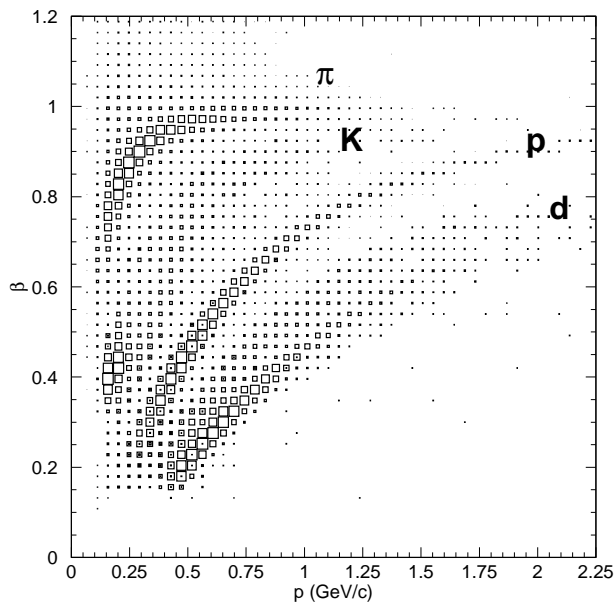}
\caption{\label{fig:beta_p} Velocity $\beta=v/c$ spectrum, as a function of particle momentum, for charged particles detected in the CLAS.}
\end{figure}

In order to isolate the $ppn$ channel, a $pp$ coincidence (with no
other charged particles) in a time window of $\pm 1$~ns with a tagged
photon defined the minimum condition for an accepted event, since the
time interval between beam pulses is 2~ns. This coincidence time is shown
in Fig.~\ref{fig:coinc_time} for a subset of the raw data.
The two protons were identified by their mass, deduced from their
momentum measured in the drift chambers and their velocity measured
with the time-of-flight scintillators, as shown in
Fig.~\ref{fig:beta_p}.

\begin{figure}
\includegraphics[scale=0.45]{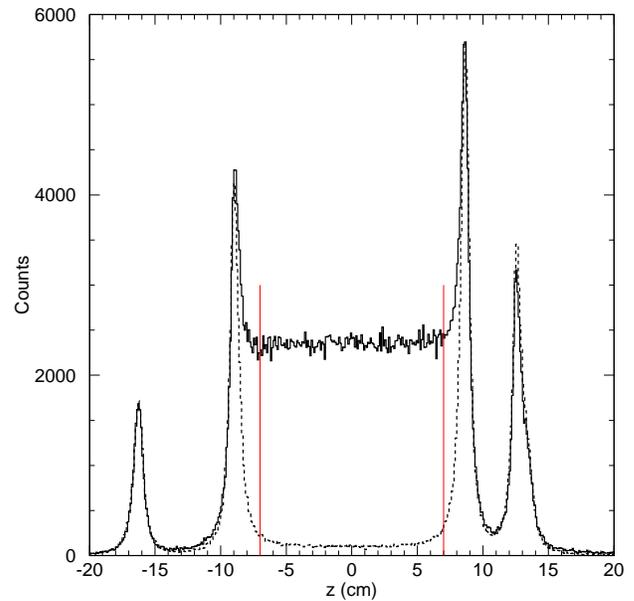}
\caption{\label{fig:z_noback} Distribution of the $z$ component (along the beam line) of the proton vertex. The solid line represents data obtained with a full target and the dashed line represents data taken with an empty target. The two inner peaks are events produced in the target walls, the two outer peaks represent protons produced in the superinsulation of the target cell and in its axial heat shield. The range $-7$ cm $ < z < 7$ cm (vertical lines) has been chosen to select the $pp$ events.}
\end{figure}

A cut on the interaction vertex, based on the analysis of empty-target runs, was performed to eliminate the background from $pp$ events originating outside the target volume. Eliminating the events having the $z$ component (where $z$ is measured along the beam line) of the vertex more than $7$ cm away from the center of the target, as shown in Fig.~\ref{fig:z_noback}, reduced this background to less than $1$\% \cite{tesi}.

\begin{figure}[htbp]
\includegraphics[scale=0.45]{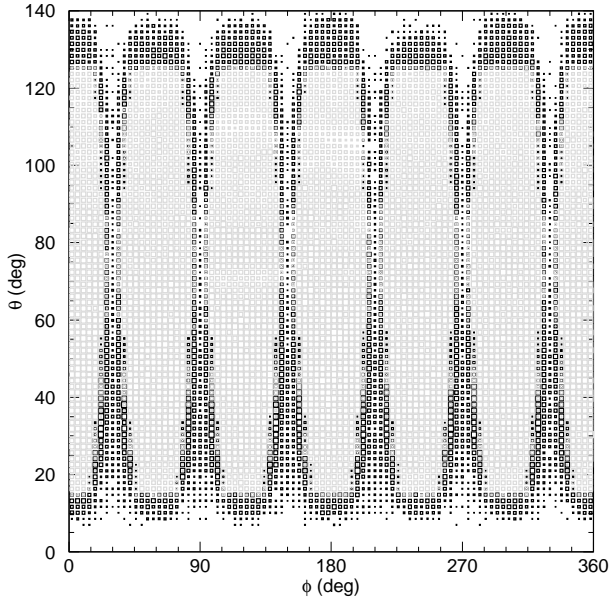}
\caption{\label{fig:fid_cuts} Angular coverage for the identified protons. The gray areas represent the fiducial regions of the six CLAS sectors inside which the protons for the present analysis have been accepted.}
\end{figure}

The particle-detection efficiency of the CLAS is not uniform and constant throughout its volume. At the edges of the active regions, delimited by the shadows of the six superconducting coils, the acceptance decreases and varies rapidly. In order to avoid errors, including poorly reconstructed tracks in the low-acceptance regions, a set of fiducial cuts, empirically determined, has been applied both to the momenta ($p_1,p_2>300$ MeV/c, $p\sim300$ MeV/c being the CLAS detection threshold for protons) and to the polar and azimuthal angles ($\theta$, $\phi$) of the protons. The requirement of having the two protons in two different sectors of CLAS has also been applied, in order to avoid inefficiencies in the reconstruction of close tracks. 
The angular coverage for the accepted protons is shown in the light grey areas of Fig.~\ref{fig:fid_cuts}. 

\begin{figure}[htbp]
\includegraphics[scale=0.45]{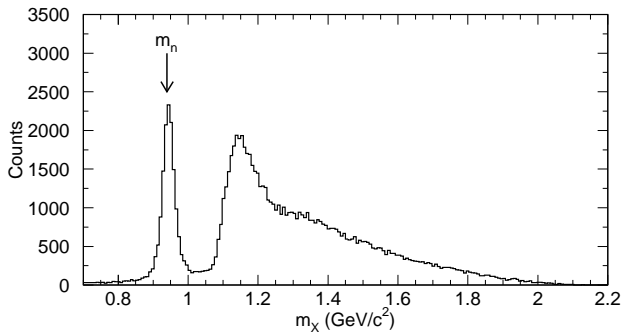}
\caption{\label{fig:missmass} Missing mass of the $\gamma^3{\mbox{He}}
\rightarrow ppX$ system, for a subset of the selected $pp$ events.
One can easily distinguish the peak at the neutron mass at about
0.94~GeV/c$^2$ ($\sigma \sim 0.017$ GeV/c$^2$) from the competing reaction channels.  }
\end{figure}

Since the photon energy and the four-momenta of the two detected
protons are known, and thus the $ppn$ kinematics is completely
determined, a missing-mass analysis can be performed to identify the
neutron.  Figure~\ref{fig:missmass} shows the missing-mass
distribution of the system $\gamma^3{\mbox{He}} \rightarrow ppX$. The
first peak corresponds to the missing neutron, the second one to the
other competing reaction channels, such as those producing pions which
had not been detected by the CLAS, {\it e.g.},
$\gamma^3{\mbox{He}}\rightarrow pp(n\pi^0)$ or
$\gamma^3{\mbox{He}}\rightarrow pp(p\pi^-)$.  About 25\% of the
two-proton events, $\sim5$ million events, are thus identified as
belonging to the $ppn$ channel.

The momentum of each detected proton was corrected for its loss of energy while passing 
through the cryogenic target material, the target walls, the carbon-fiber scattering chamber and the start-counter scintillators.

\subsection{\label{sec:background}Background Subtraction}
\begin{figure}
\includegraphics[scale=0.45]{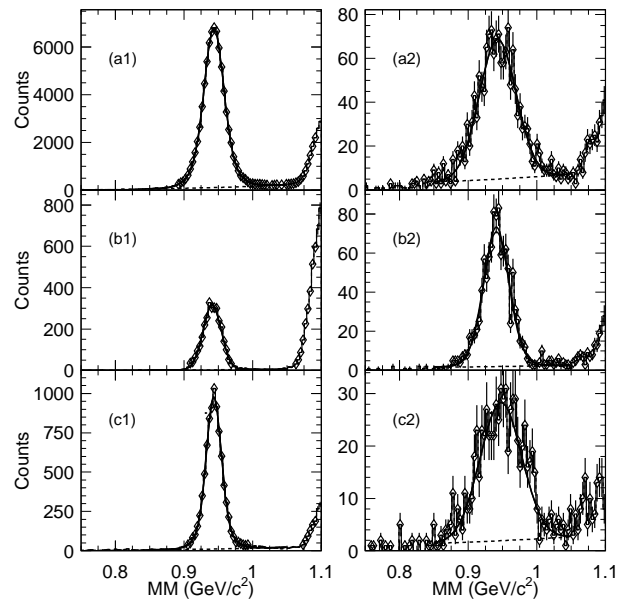}
\caption{\label{fig:fit_exp} Examples of missing-mass histograms fitted with a Gaussian curve plus an exponential (solid curve) for $0.43<E_{\gamma}<0.45$ GeV (a1) and $1.13<E_{\gamma}<1.15$ GeV (a2), for $0.08<p_n<0.10$ GeV/c and $0.45<E_{\gamma}<0.55$ GeV (b1) and $0.42<p_n<0.44$ GeV/c and $0.75<E_{\gamma}<0.85$ GeV (b2), and for $-0.88<\cos\theta_n<-0.84$ and $0.35<E_{\gamma}<0.45$ GeV (c1) and $0.72<\cos\theta_n<0.76$ and $0.95<E_{\gamma}<1.05$ GeV (c2). The background alone is shown as the dashed curves.}
\end{figure}

After channel identification, the data were binned in photon energy,
particle momentum, and particle angle. For each of these bins a
histogram of the two-proton missing-mass distribution was
accumulated. 
Each $pp$ missing-mass histogram was fitted with a Gaussian curve plus
an exponential in order to reproduce the neutron peak and the
background underneath it.  The background is due both to misidentified
or badly reconstructed protons and to the tail from competing reaction
channels (see Fig.~\ref{fig:missmass}). Once the parameters of the fit
are extracted, the yield is given by the area under the Gaussian
curve. In this way, the contribution of the background is excluded.
Some examples of the quality of these fits for various bins in photon energy,
neutron momentum, and neutron angle, chosen to be typical of
the character of the data for various conditions, are shown in
Fig.~\ref{fig:fit_exp}. The background-to-signal ratio varies from
less than $1\%$ to $8\%$, depending on the kinematics.

To estimate the systematic uncertainty introduced by the fitting
procedure used to subtract the background from the $pp$ missing mass,
the yields obtained with two kinds of fitting functions for the
background (exponential and polynomial) have been compared with each
other \cite{tesi}. The deviations are, on average, of the order of
2\%.

\subsection{\label{sec:efficiency}Efficiency}
Since the neutron is reconstructed using the missing-mass technique,
the detection efficiency for this channel is given by the probability
of correctly detecting and identifying two protons in the CLAS. This
has been evaluated with the aid of a Monte-Carlo simulation. The $ppn$
events, generated according to the three-body phase-space
distribution, were processed by a GEANT-based code simulating the
response of the CLAS, and were reconstructed and analyzed using the
same procedure adopted for the experimental data. 
The efficiency inside the CLAS fiducial region for a given kinematical bin
$\Delta\tau$ is defined as
\begin{eqnarray}\label{eq:efficiency}
\varepsilon(\Delta\tau)=\frac {N_r}{N_0}\;,
\end{eqnarray}
where $\Delta\tau$ lies inside of the CLAS fiducial region, $N_r$ is
the number of reconstructed events within $\Delta\tau$, and $N_0$ is the
number of events generated within $\Delta\tau$.
The efficiency so computed is more or less constant as a function of photon energy, momentum, and angles, and its average value is slightly less than 95\%.

In order to evaluate the systematic uncertainties in the efficiency
for detecting the $ppn$ events in the CLAS, the results obtained with
the phase-space distribution have been compared with the efficiency
computed with three other event distributions \cite{tesi}.  The result
of the calculations of the efficiency inside of the CLAS fiducial
region turns out to be independent of the model used to simulate the
reaction, apart from the effect of bin migration due to the finite
resolution of the detector, which has been found to be small. The
resulting systematic uncertainty was determined to be no greater than
5\% \cite{tesi}.

\subsection{\label{sec:normalization}Cross Sections and Normalization}
Three kinds of CLAS-integrated cross sections have been measured and are reported here. They are
\begin{itemize}
\item{total cross sections, defined as
\begin{eqnarray}\label{eq:tot_cross_sec}
\sigma=\frac{N_{ppn}}{L\varepsilon}
\end{eqnarray}}
\item{semi-differential cross sections with respect to momentum, defined as
\begin{eqnarray}\label{eq:diff_p_cross_sec}
\frac {d\sigma}{dp}=\frac{N_{ppn}}{\Delta p L \varepsilon}
\end{eqnarray}}
\item{semi-differential cross sections with respect to $\cos\theta$, defined as
\begin{eqnarray}\label{eq:diff_th_cross_sec}
\frac {d\sigma}{d\Omega}=\frac{N_{ppn}}{2\pi\Delta (\cos\theta) L \varepsilon}\;,
\end{eqnarray}}
\end{itemize}
where $N_{ppn}$ is the number of events in the bin, $\varepsilon$ is
the detection efficiency defined in Section~\ref{sec:efficiency}, and
$L$ is the luminosity, which is defined as:
\begin{eqnarray}\label{eq:luminosity}
L=N_{\gamma}\frac{\rho z N_A}{A}\;,
\end{eqnarray}
where $\rho=0.0675$~g/cm$^3$ is the density of the target, $z=14.0$~cm
is the effective target length, $A$ is the atomic mass of the target
($A=3.016$ g/mol), $N_A$ is Avogadro's number, and $N_{\gamma}$ is the
number of incident photons.

\begin{figure}
\includegraphics[scale=0.45]{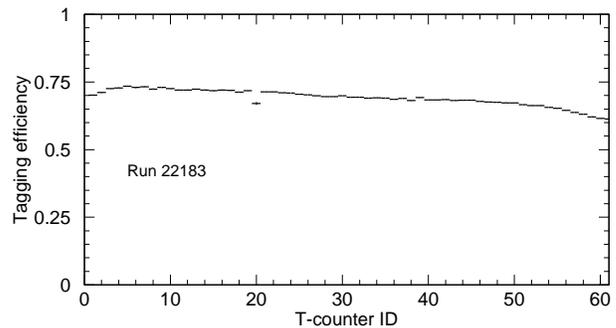}
\caption{\label{fig:tag_eff} Tagging efficiency as a function of T-counter number measured in one particular low-flux run. The average efficiency is about 70\%.}
\end{figure}

The systematic uncertainties in the target length and density are of the order of 2\%.
The photon flux was measured by integrating the tagger rate over the data-acquisition lifetime. The tagging efficiency was measured during low-flux runs, using the lead-glass total absorption detector. For each T-counter $i$, the tagging efficiency is defined as \cite{tagger}:
\begin{eqnarray}\label{eq:tag_eff}
T_{eff}(i)=(T_i \cdot TAC)/T_i^{raw}\;,
\end{eqnarray}
where $T_i \cdot TAC$ is the rate of coincidences between tagger and total absorption counter, and $T_i^{raw}$ is the rate in the tagger alone.
A typical tagging efficiency spectrum, as a function of T-Counter number, is shown in Fig.~\ref{fig:tag_eff}.
To estimate the systematic uncertainty for the photon flux, the
variations with time of the tagging efficiency and of the 
proton yield normalized to the photon flux for each tagger 
scintillation counter have been
studied. The resulting systematic uncertainty is, on average,
approximately 6\% \cite{tesi}.  The values of the systematic
uncertainties in the measured cross sections are summarized in
Table~\ref{table:systematic}.  The luminosity, integrated over the
entire running time and over the full photon-energy range, was
$L\simeq8.7\times 10^{35}$ cm$^{-2}$ for this experiment.

\begin{table}[ht]
\caption{\label{table:systematic} Systematic uncertainties in the measured cross sections. The total is the sum in quadrature of the individual uncertainties.}
\begin{tabular}{ c  c }
\hline\hline
{Quantity} & {Uncertainty}\\
\hline
{Target length and density} & {$2\%$}\\
{Background subtraction} & {2\%}\\
{Detection efficiency} & {5\%}\\
{Photon flux} & {6\%}\\
\hline
{Total} & {$8\%$}\\
\hline\hline
\end{tabular}
\end{table}

\section{\label{sec:res_full}Model calculation}

As mentioned in the introduction, the only theoretical model currently
available for calculation of the cross section for the three-body
photodisintegration of $^3$He in the GeV energy region is the one by
Laget. 
In this model, the five-fold differential cross section in the laboratory system for the $\gamma^3{\mbox{He}} \rightarrow ppn$ reaction is connected through a Jacobian to a {\it reduced cross section} 
\begin{eqnarray}\label{eq:laget}
\frac{d^5\sigma}{dpd\Omega_1d\Omega_2}& = &\frac{E_np_2^3p_1^2}{E_1 p^2_n|E_np_2^2-E_2 \overrightarrow{p_n} \cdot \overrightarrow{p_2}|}\left(\frac{Q}{p}\right)_{cm} \cdot \nonumber\\
                                      & \cdot &  \frac{d^5\sigma_{red}}{(d{\Omega_1})_{cm}dp_n d\Omega_n}
\end{eqnarray} 
where $(E_1,\overrightarrow{p_1})$, $(E_2,\overrightarrow{p_2})$, and $(E_n,\overrightarrow{p_n})$ are, respectively, the four-momenta of the two outgoing protons (1 and 2) and the neutron in the laboratory frame, and $p$ and $Q$ are the proton momentum and the total energy measured in the center-of-mass frame of the two protons.

\begin{figure}
\includegraphics[scale=0.35]{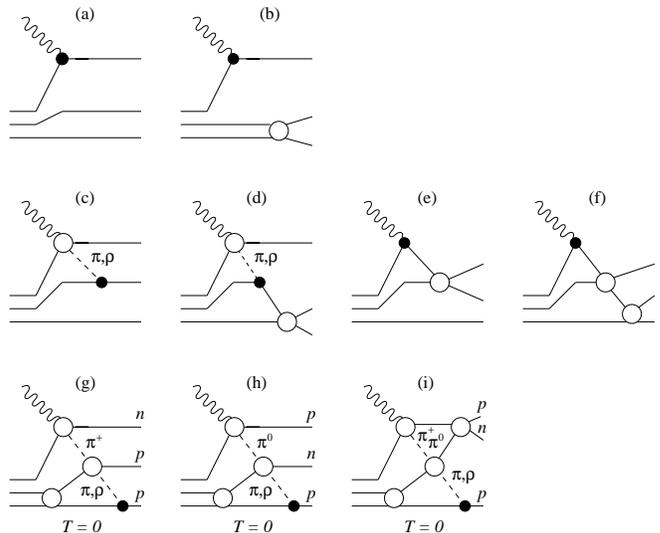}
\caption{\label{fig:diagrams} Diagrams used in Laget's model \cite{laget_a, laget_b, laget_c, laget_d, laget_e} in the calculation of the $^3$He$(\gamma,pp)n$ cross sections: (a) 1N absorption mechanism, (b) 1N+Final State Interactions (FSI), (c) 2N absorption, (d), (e) and (f) 2N + FSI, (g) and (h) 3N mechanisms, and (i) 3N+FSI. The open circles represent full transition amplitudes (T matrices); the filled circles are $\gamma NN$ and $\pi NN$ Born terms.}
\end{figure}

The reduced cross section depends on the transition amplitude T$(\gamma ^3He\rightarrow ppn)$ \cite{laget1,laget1b}:
\begin{eqnarray}\label{eq:diff_cross_sect}
\frac{d^5\sigma_{red}}{(d\Omega_1)_{cm} dp_n d\Omega_n}\propto |\langle \Psi_{ppn}|T|\Psi_{^3He}\rangle|^2.
\end{eqnarray}
The fully antisymmetrized $^3$He bound-state wave function
$|\Psi_{^3He}\rangle$ is the solution of the Faddeev equations
\cite{hajduk81} for the Paris potential \cite{lacombe80}. It is
expanded in a basis where two nucleons couple to angular momentum $L$,
spin $S$, and isospin $T$, the third nucleon moving with angular
momentum $l$. Each component is approximated by the product of the
wave functions, which describe the relative motion of the two nucleons
inside the pair and the motion of the third nucleon
\cite{laget2}. Fermi-motion effects are taken fully into account in
the two-body matrix element, and partially \cite{laget_d} in the
three-body matrix element. However, it has been ascertained that the
effect of the Fermi motion in the three-body matrix element does not
significantly affect the results; therefore, it has not been
implemented in the version of the model which has been used here with
the Monte-Carlo procedure in order to avoid prohibitive computation
time.  All of the S, P, and D components of the $^3$He wave function
are included.  The energy and momentum are conserved at each vertex,
and the kinematics is relativistic.  The continuum final state
$|\Psi_{ppn}\rangle$ is approximated by a sum of three-body plane
waves and half-off-shell amplitudes (which are the solutions of the
Lippman-Schwinger equation for the Paris potential) where two nucleons
scatter, the third being a spectator. Only S-wave NN scattering
amplitudes have been retained in the version used in this work. The
antisymmetry of the continuum final state is achieved by exchanging
the roles of the three nucleons \cite{laget_c}.  The transition
amplitude $T$ is expanded in a truncated series of diagrams that are
thought to be dominant.  These diagrams, that were thought to include
the most likely one-, two- and three-body mechanisms, are computed in
momentum space.  Among all the possible three-body mechanisms, meson
double scattering is the most likely to occur. The Feynman diagrams
included in the present version of the model are shown in
Fig.~\ref{fig:diagrams}.  The open circles represent the full
transition amplitudes (T matrices), which have been calibrated against
the corresponding elementary channels, and the filled circles are just
the $\gamma NN$ and $\pi NN$ Born terms.

The first two diagrams, (a) and (b), describe one-body
photoabsorption; (c), (d), (e), and (f) represent two-body processes
\cite{laget3_a, laget3_b}; and (g), (h), and (i) are three-body
mechanisms, with two-meson ($\pi$ or $\rho$) exchange. Pion absorption
by a $T=1$ ($pn$ or $pp$) pair has been found experimentally to be
strongly suppressed \cite{aniol86}, at least at low energies, and has
not been included in the model at this stage.  The $3N$ absorption
mechanism shown in diagram (g) represents the primary $3N$ process for
the $^3$He$(\gamma,pp)n$ reaction. Above the photon energy
corresponding to the pion-production threshold, the calculation does
not contain any free parameters, since all of the basic matrix
elements have been fixed independently using relevant reactions
induced on the nucleon and on the deuteron \cite{laget1,laget1b}. The
calculated cross section involves a logarithmic singularity associated
with the on-shell propagation of the ``first'' exchanged pion, which
shows up, and moves when the photon energy varies, in a well defined
part of the phase space. Below the pion threshold, both exchanged
pions are off their mass shells, and the three-body exchange currents
can be linked by gauge invariance to the corresponding three-body
forces \cite{laget_a, laget_b, laget_c, laget_d, laget_e}.

All model calculations discussed in the following sections have been
performed with Monte-Carlo sampling over the CLAS geometry to produce
cross sections that can be compared with the experimental results. The
small-scale structures which are seen in some of the model results
result from this Monte Carlo treatment, although the major structures
are real features of the model calculations.

\section{\label{results}RESULTS}

\subsection{Cross Sections Integrated over CLAS}

The use of a triangular Dalitz plot is very suitable to look for the deviations of an experimental distribution from pure phase-space predictions and to identify correlations between three final-state particles. 
In particular, this technique can be used to identify and select the regions of the phase space where three-body processes can be dominant. 
If $T_{p1}$, $T_{p2}$, and $T_n$ are the center-of-mass kinetic energies of the two protons and the neutron, respectively, and $T$ is their sum, we can define the Cartesian coordinates of the triangular Dalitz plot as:
\begin{eqnarray}\label{dalitz_coord}
x & = & \frac{1}{\sqrt 3}\frac{T_{p1}-T_{p2}}{T}\;\;\; {\rm and}\nonumber \\
y & = & \frac {T_n}{T}\nonumber.
\end{eqnarray}

\begin{figure}[htbp]
\includegraphics[scale=0.4]{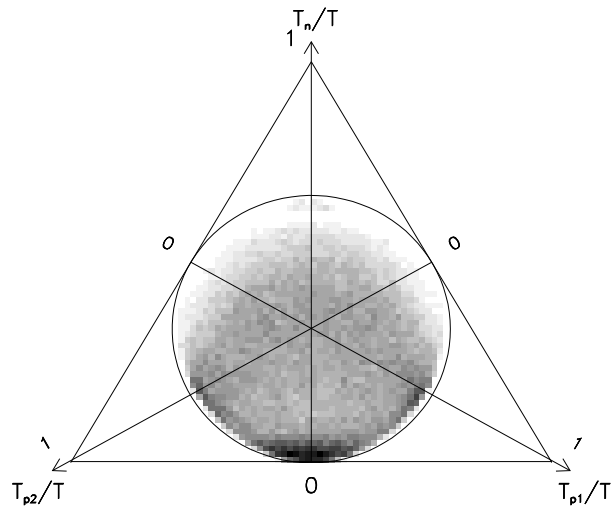}
\caption{\label{fig:dalitz_data} Triangular Dalitz plot for the $ppn$ data. $T_{p1}$, $T_{p2}$, and $T_{n}$ are the center-of-mass kinetic energies of the three nucleons.}
\end{figure}

Figure~\ref{fig:dalitz_data} shows the distribution of the $ppn$
events on the Dalitz plot after applying the selection cuts.  The wide
acceptance of the CLAS allows us to fill the physically accessible
region --- delimited by the boundary circle --- almost completely.

The shading of the boxes indicate the yield of the observed $ppn$
events. Areas of increased yield are visible where the $T_{p1}$ and
$T_{p2}$ axes intercept the boundary circle, as well as where $T_n
\approx 0$. These areas correspond to quasi-two-body breakup and
neutron-spectator kinematics, respectively; they are discussed in
detail in Sections \ref{sec:side} and \ref{sec:res_spectator} below.
The depletion areas in the upper left and upper right sides of the circle
correspond to the kinematics where one of the protons has low
momentum ($p_1,p_2<300$ MeV/c) and therefore is not detected by the
CLAS. The central top area where the two protons are emitted in nearly
the same direction is excluded by the requirement of detecting the two
protons in two different sectors (see Section~\ref{sec:ch_id} above).
The central region, near the intersection of the three axes,
consists of events where all three nucleons have nearly equal
energies, and is called the ``star'' region (see Section
\ref{sec:res_star} below).

In the following sections, CLAS-integrated cross sections for the
full acceptance and for the three selected kinematics listed above,
each chosen to illustrate its two-body or three-body character, are
presented and compared with distributions obtained both with
three-body phase space and with the results of the Laget model.

\subsubsection{Full CLAS acceptance}

\begin{figure}
\includegraphics[scale=0.48]{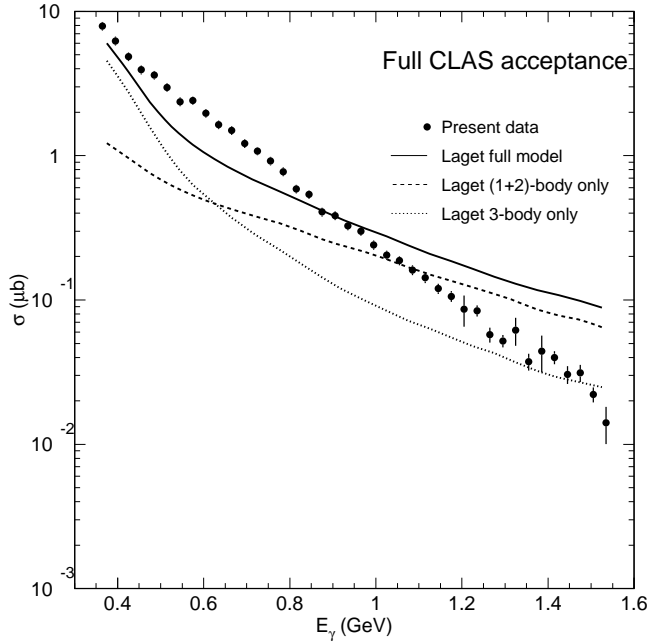}
\caption{\label{fig:egamma_nocuts} Total $ppn$ cross section
integrated over the CLAS acceptance plotted as a function of photon
energy on a logarithmic scale for the full $E_{\gamma}$ range. The
$ppn$ cross section (circles) is compared with Laget's full model
(solid curve), with the model result without the three-body mechanisms
(dashed curve), and with the one including only three-body processes
(dotted curve). The error bars include statistical and systematic
uncertainties, as in all the following experimental distributions.}
\end{figure}

The $ppn$ total cross section integrated over the CLAS acceptance has been measured as a function of the incident photon energy $E_{\gamma}$. 
The photon-energy spectrum, ranging from 0.35 GeV to 1.55 GeV, has been divided into 60 bins, each 0.02 GeV wide.
The results are shown in Fig.~\ref{fig:egamma_nocuts}. The cross section, ranging between $10\; \mu$b and $0.01\;\mu$b, decreases almost exponentially as the photon energy increases. Fitting the data with an exponential function $\sigma(E_{\gamma})\propto e^{-bE_{\gamma}}$ yields a slope $b\simeq5.3$ GeV$^{-1}$.
The data are compared with the full calculation (solid curve),
including one-, two-, and three-body mechanisms, as well as with the
results for the one- and two-body mechanisms only (dashed curve), and
the three-body mechanisms only (dotted curve), as shown in
Fig.~\ref{fig:diagrams}. It is important to note that the theoretical
curves represent {\it absolute cross sections} calculated within the
CLAS acceptance -- they are not normalized to the data. The results of
the model calculations that do not include the three-body mechanisms
are almost a factor of ten smaller than the data at lower energies,
while they approach the data as the photon energy increases and exceed
the data at higher energies. The full-model results agree better with
the data, but still are too low at low energies and too high at high
energies.

\begin{figure*}
\includegraphics[scale=0.5]{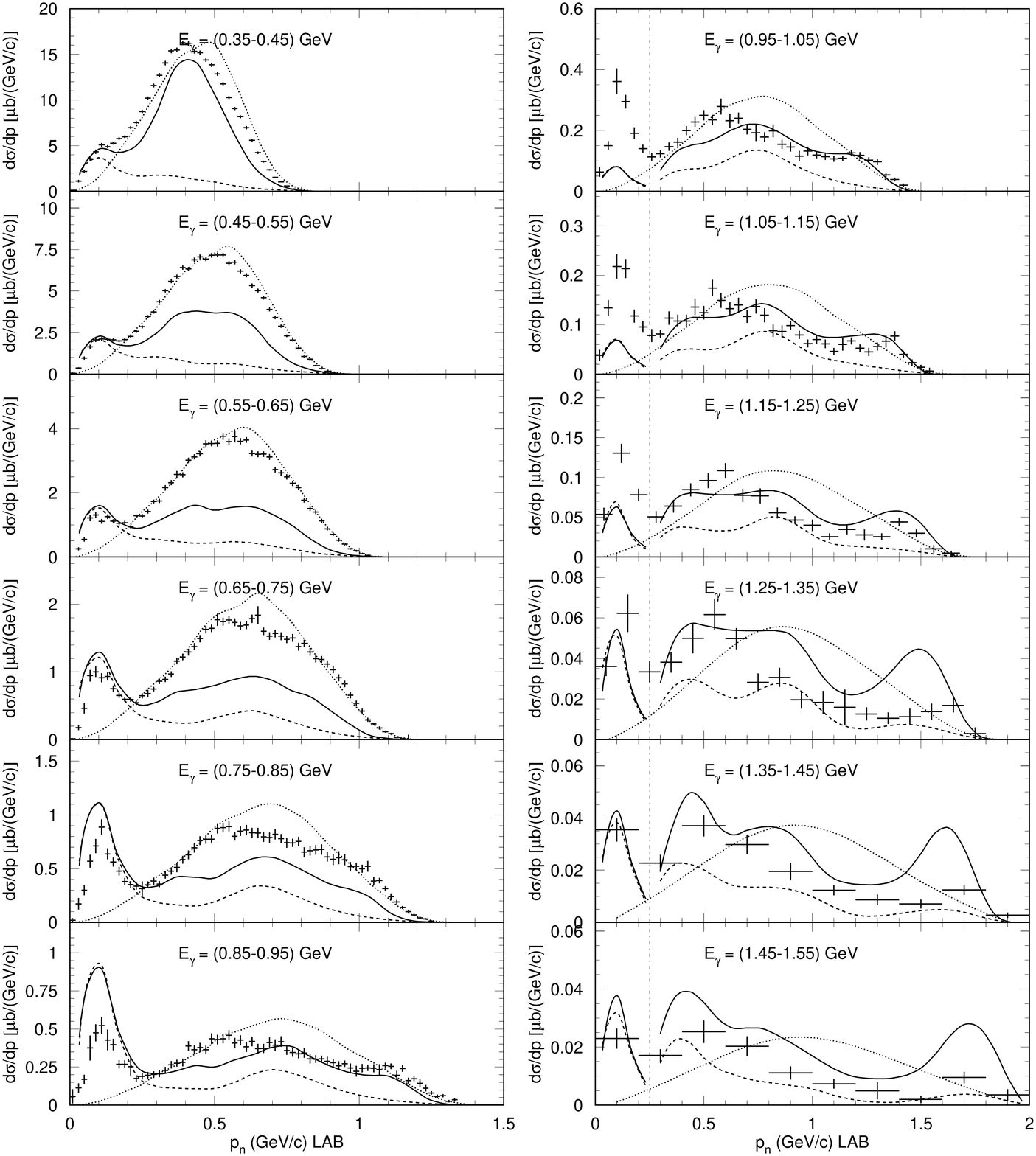}
\caption{\label{fig:pn_nocut} Differential cross sections
integrated over the CLAS as a function of the neutron momentum in the
laboratory frame
for twelve 0.1-GeV-wide photon-energy bins between 0.35 GeV and
1.55 GeV. The points represent our CLAS data. The error bars include
both statistical and systematic uncertainties. The dotted curves are
the distributions for phase-space events generated within the CLAS
acceptance and normalized in each energy bin to match the total area
of each experimental distribution. The solid curves represent the full
Laget-model results, while the dashed lines represent the model
including one- and two-body mechanisms only. For $E_{\gamma}>0.95$
GeV, the model predictions at $p_n<250$ MeV/c (to the left of the
vertical dotted-dashed line) are scaled by a factor 0.1 to fit in the
plots.}
\end{figure*}

Figure~\ref{fig:pn_nocut} shows the partial differential cross section
as a function of neutron momentum $p_n$, for twelve 0.1-GeV-wide
photon energy bins. The data are compared with phase-space-generated
event distributions (dotted curves) normalized in each energy bin in
order to match the area under the experimental distribution, with the
results of Laget's full model (solid curve), and with the model with no
three-body mechanisms included (dashed curve). The neutron
momentum distributions are related to the projection of the data in
Fig.~\ref{fig:dalitz_data} onto the $T_n$ axis.

In the photon-energy range between 0.35 and 0.95 GeV, the data show a
broad central distribution in the middle of the neutron momentum
spectrum ({\it e.g.}, at about 400 MeV/c for $E_{\gamma}=0.4$ GeV and
500 MeV/c for $E_{\gamma}>0.5$ GeV), which is reproduced reasonably
well by the phase-space distribution (better at low photon energies
than at high energies). Up to about 0.6~GeV a comparison of the data
with the shape of the model results reveals the presence of three-body
mechanisms. In the middle range of neutron momentum, two-body
mechanisms are seen to contribute increasingly starting from
$E_\gamma=0.65$ GeV. These contributions stem from low-energy S-wave
$np$ rescattering, which causes the increased yield in the
quasi-two-body kinematics, corresponding to the the left and right 
sides of the Dalitz plot
(Fig.~\ref{fig:dalitz_data}). This yield projects onto the middle
range of the neutron-momentum distribution.

A peak, roughly 0.04-GeV/c wide, is observed at a neutron momentum of
about 0.12 GeV/c, independent of the photon energy. The relative
strength of this peak increases with increasing photon energy, but it is not
accounted for by the three-body phase-space distribution. However,
this structure is expected by the model, and it is predicted to be
largely due to two-body mechanisms. It reflects the Fermi distribution
of the spectator neutron. 
This feature has been exploited to select the neutron-spectator
kinematic region, as is explained in Section
\ref{sec:res_spectator} below.
 
At photon energies from about 0.9 to 1.2~GeV and high neutron momenta,
a third structure appears in the data, which is present neither in the
phase-space distribution nor in the (1+2)-body model results, but is
predicted by the full model. This structure can therefore be
considered to be a signature of three-body mechanisms as well.

\begin{figure*}
\includegraphics[scale=0.5]{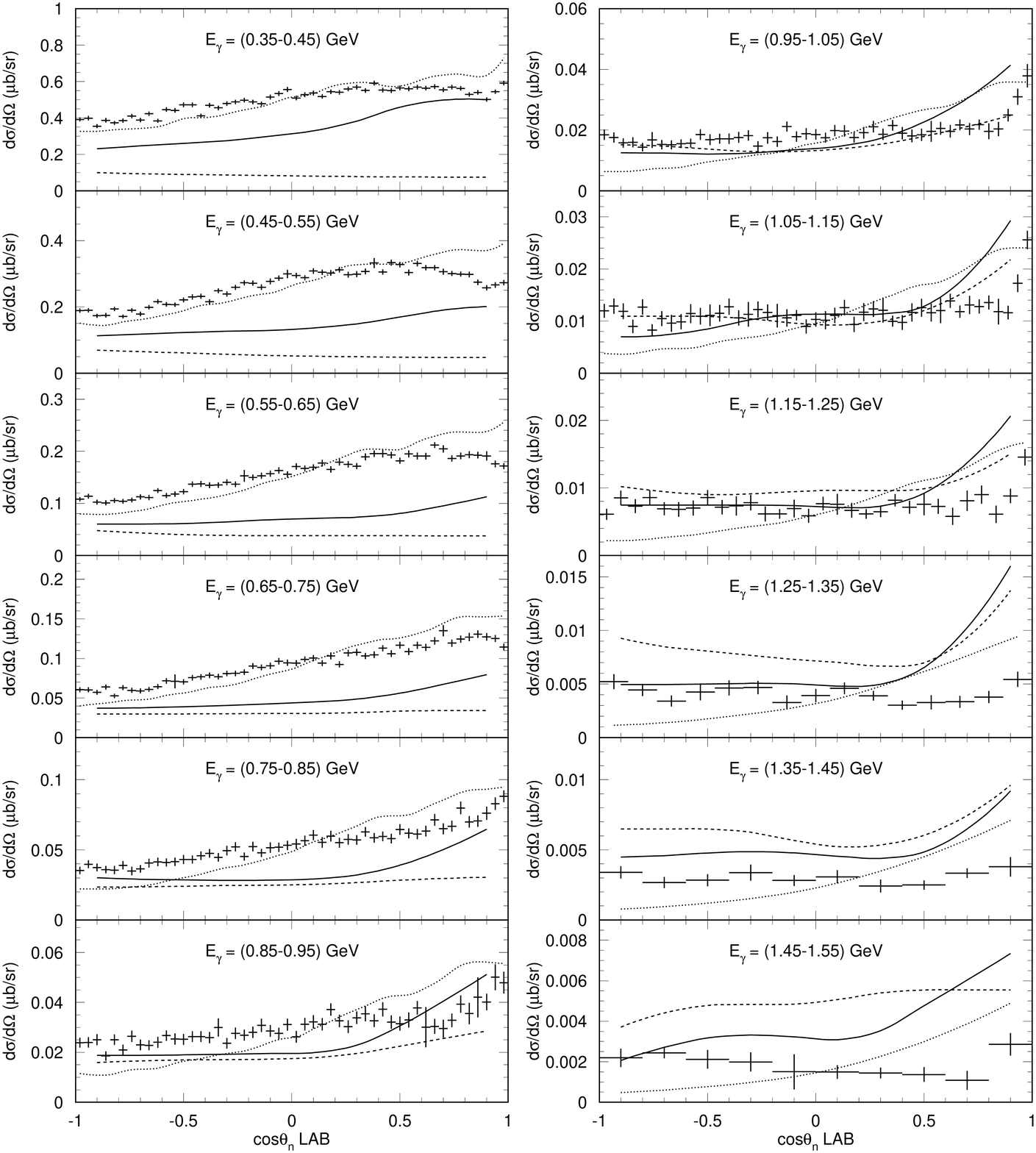}
\caption{\label{fig:thn_nocut} Differential cross sections integrated over the CLAS as a function of the cosine of the neutron polar angle in the laboratory frame
for twelve 0.1-GeV-wide photon-energy bins between 0.35 GeV and 1.55 GeV. The points represent our CLAS data. The
error bars include statistical and systematic uncertainties. The
dotted curves are the distributions for phase-space events generated
within the CLAS acceptance and normalized in each energy bin to match
the total area of each experimental distribution. The solid curves
represent the full model results, while the dashed lines represent the
model including one- and two-body mechanisms only.}
\end{figure*}

The differential cross section as a function of the cosine of the
neutron polar angle $\cos\theta_n$ in the lab system is plotted in
Fig.~\ref{fig:thn_nocut}, for twelve 0.1-GeV-wide photon-energy bins,
between 0.35 GeV and 1.55 GeV.  The distributions are forward-peaked
at low-to-intermediate energies, while they become flatter for higher
$E_{\gamma}$. Their shapes are reasonably well reproduced by both
phase-space and the full-model calculations.

\subsubsection{\label{sec:res_spectator}Spectator neutron}

Guided by Fig.~\ref{fig:pn_nocut}, the events where the neutron is a spectator in the photobreakup of a proton pair have been selected by requiring the condition $p_n<250\; {\rm MeV/c}$.
These are all the events in the lower neutron-momentum peak (within $3\sigma$ from its center). 

\begin{figure}
\includegraphics[scale=0.48]{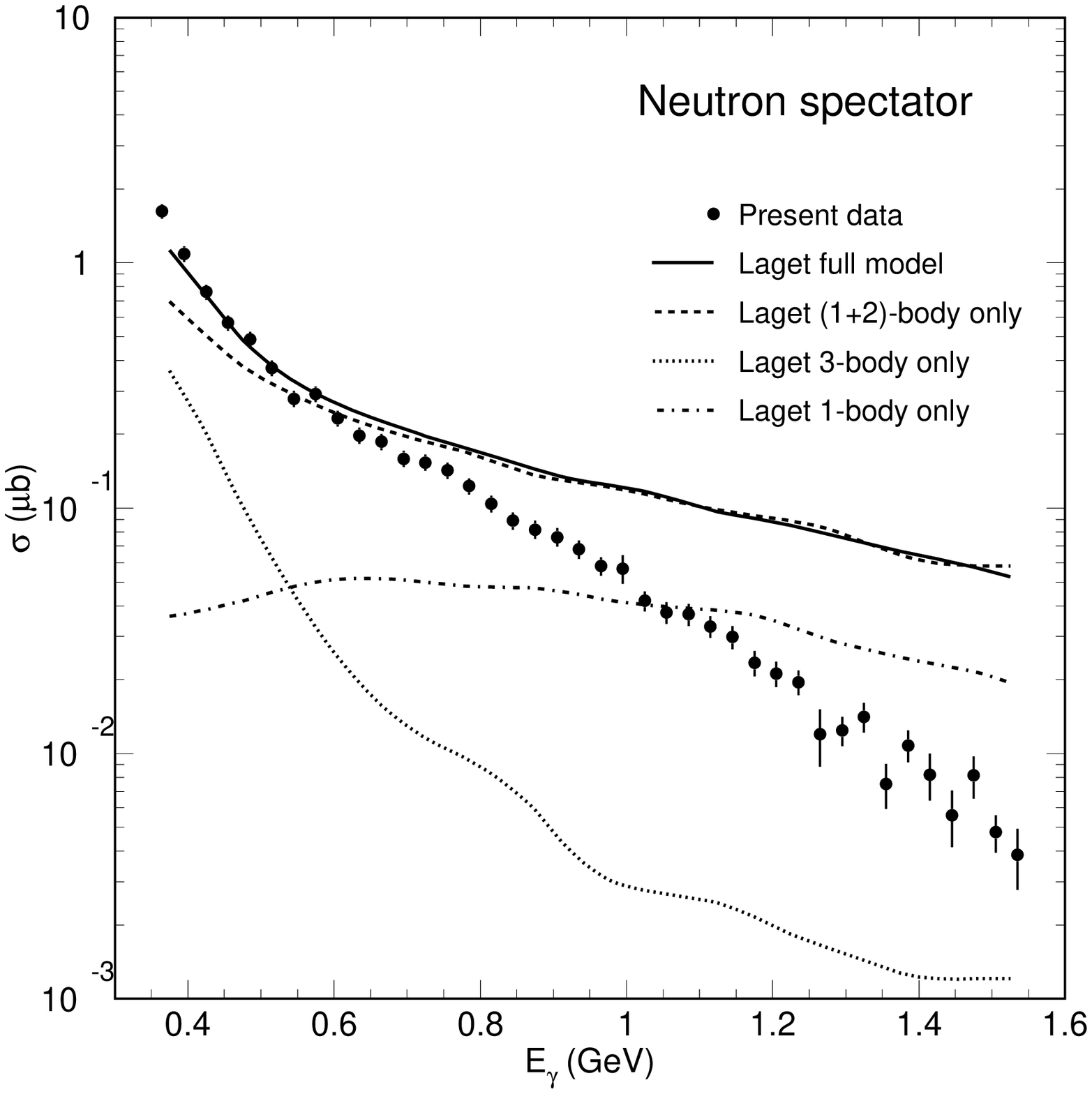}
\caption{\label{fig:spectator_egamma} Cross section integrated over the CLAS for the neutron-spectator kinematics plotted as a function of photon energy. The data are compared with the predictions of the full model (solid curve), the (1+2)-body part (dashed curve), the three-body part (dotted curve), and the 1-body part alone (dashed-dotted curve).}
\end{figure}

Figure~\ref{fig:spectator_egamma} shows the cross section as a function of
photon energy integrated
over the CLAS for the events satisfying this condition, compared with
the predictions of the model. After an initial steep drop, the cross
section has an exponential dependence on the photon energy above 0.6
GeV, this time with a slope $b\simeq4$ GeV$^{-1}$. The agreement
between the experimental cross section and the model prediction is
good only for low energies, below 600 MeV. The cross section is clearly
driven by two-body mechanisms, as expected.

\begin{figure*}
\includegraphics[scale=0.5]{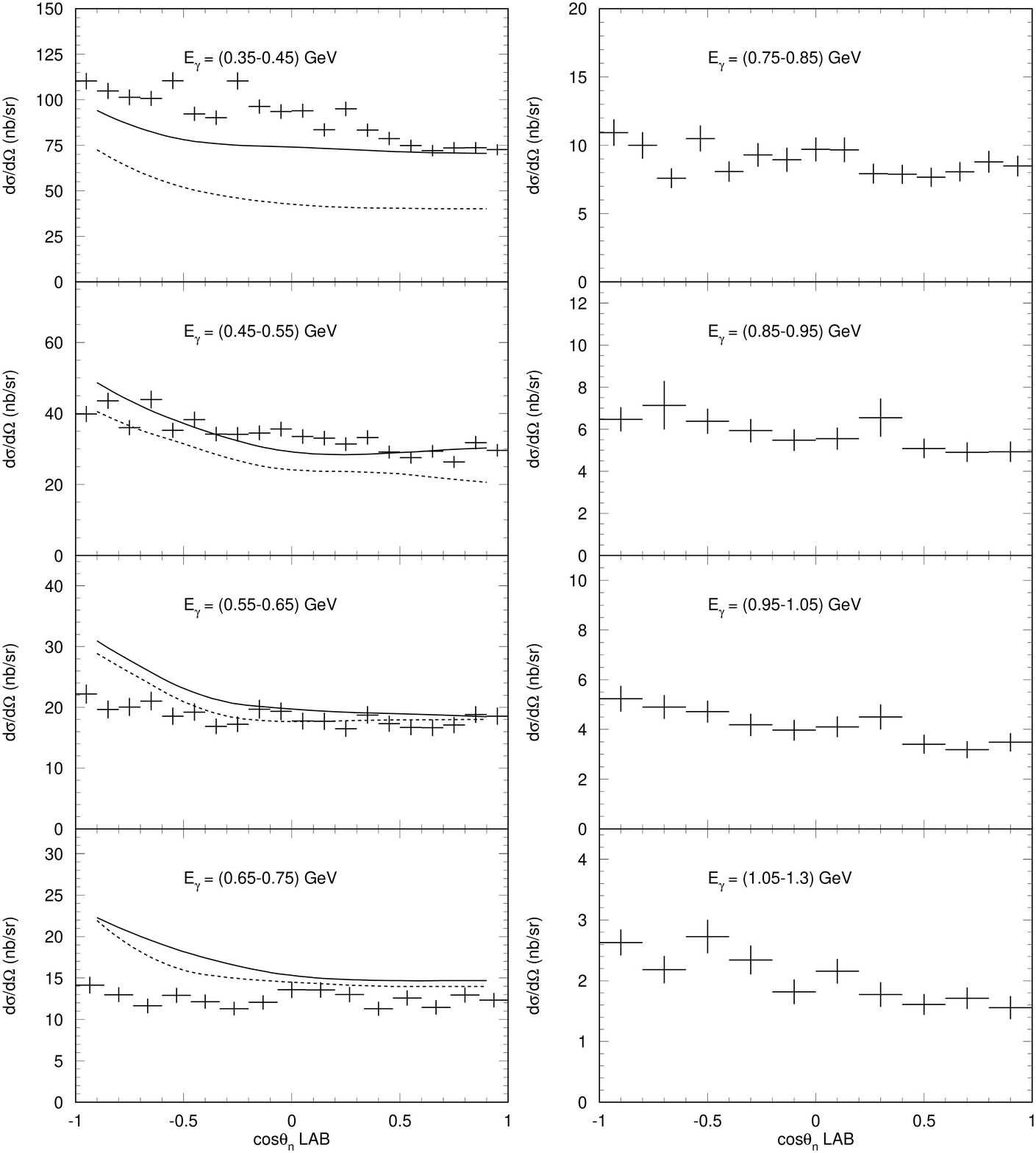}
\caption{\label{fig:spectator_th_1} Differential cross sections
integrated over the CLAS for the neutron-spectator kinematics with
respect to $\cos\theta$ of the neutron in the laboratory frame for
photon energies between 0.35 and 1.30 GeV. The data are compared with
the results of the full model (solid curves) and those of the one-
plus two-body-only model (dashed curves), for $0.35<E_{\gamma}<0.75$
GeV only, because at higher energies the model calculations differ by
more than an order of magnitude from the data.}
\end{figure*}

The differential cross section as a function of
$(\cos\theta_n)_{lab}$, which is plotted in
Fig.~\ref{fig:spectator_th_1} for eight photon-energy bins, shows a
generally flat distribution.  This is expected, because in the
neutron-spectator kinematics the two-body part of the reduced
differential cross section is proportional to the neutron-momentum
distribution $\rho(n)$ times the center-of-mass differential cross
section for the $pp$-pair breakup \cite{laget_a}:
\begin{eqnarray}\label{pp_pair}
\frac{d\sigma_{red}}{d\Omega_{cm}d\overrightarrow p_n}=(1+\beta_n \cos\theta_n) \rho(p_n) \frac{d\sigma}{d\Omega_{cm}}(\gamma pp \to pp).
\end{eqnarray}

Both the (1+2)-body part and the full-model results agree fairly well,
in shape and magnitude, with the experimental distributions up to 600
MeV. At higher energies, the calculation predicts the contribution of
two-body mechanisms to be much too large.

\begin{figure*}
\includegraphics[scale=0.5]{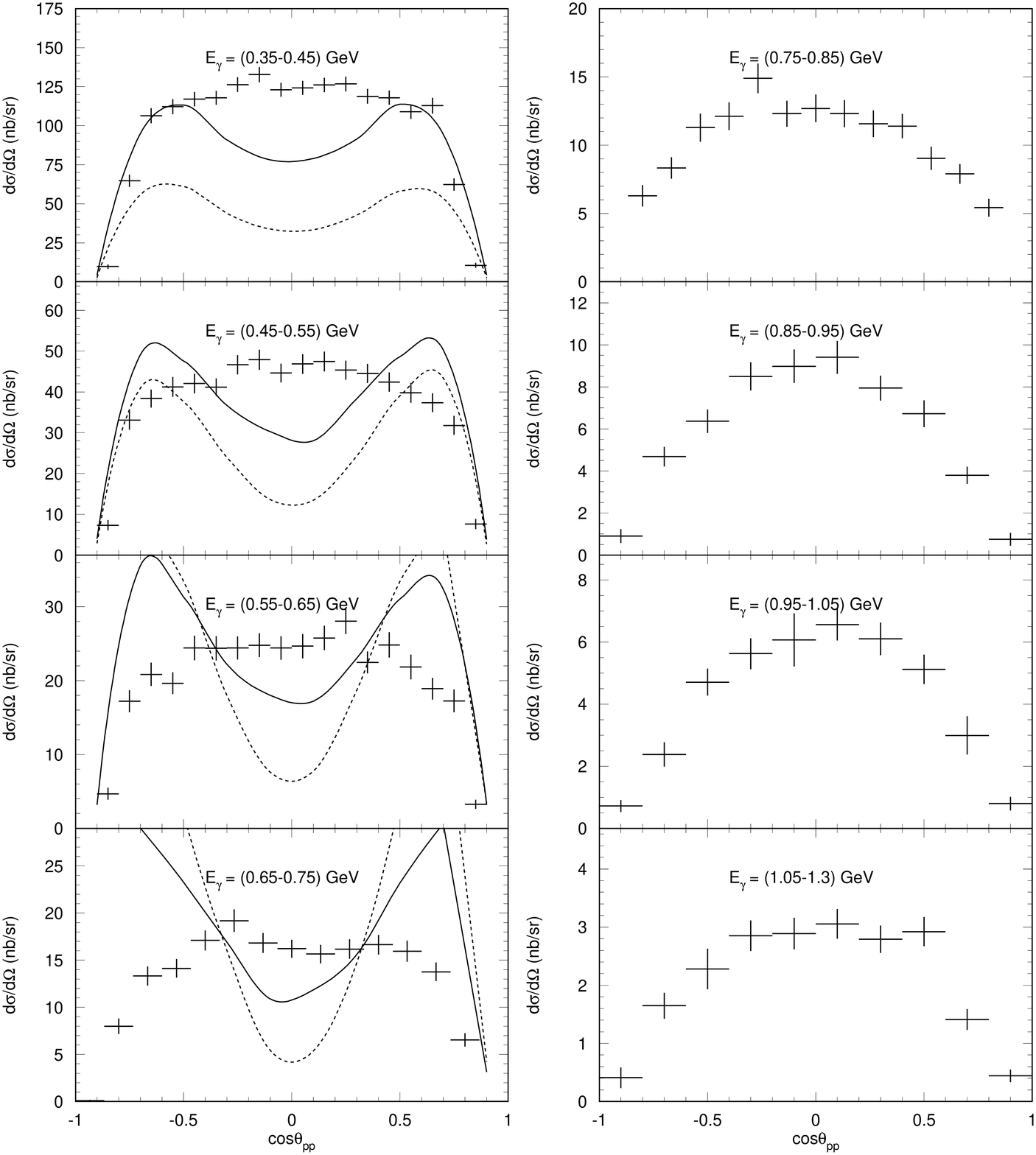}
\caption{\label{fig:spectator_th_p} Differential cross sections
integrated over the CLAS for the neutron-spectator kinematics with
respect to $\cos\theta_{pp}$ of the proton in the $pp$-pair
center-of-mass frame for photon energies between 0.35 and 1.30
GeV. The data are compared with the results of the full model (solid
curves), and those of the one- plus two-body-only model (dashed
curves), for $0.35<E_{\gamma}<0.75$ GeV, because at higher energies
the model calculations differ by more than an order of magnitude from
the data.}
\end{figure*}

In the neutron-spectator kinematics, the primary physics is contained
in the angular distribution of the $\gamma pp \to pp$
subchannel. Figure~\ref{fig:spectator_th_p} compares this angular
distribution with the model.  While the magnitude of the experimental
cross section is well reproduced at low energy by the model, the shape
of the angular distribution is markedly different. The model curve
exhibits a minimum at $90^{\circ}$, where the measured differential
cross section has a broad maximum. 
It can be seen from Fig.~\ref{fig:spectator_egamma} that three-body
diagrams do not contribute significantly to the total cross section,
but their interference with the two-body diagrams brings the shape of
the angular distributions closer to the experimental ones. However,
this effect is not strong enough to cancel the huge contribution of
the two-body part at high energy.

Since the $pp$ pair that absorbs the photon has no dipole moment for
the photon to couple with, charged-meson exchange currents and
intermediate-$\Delta$ production (Fig.~\ref{fig:diagrams}, diagrams
(c) and (d)) are strongly suppressed and one-body mechanisms (diagrams
(a) and (b)) and related FSI (diagrams (e) and (f)) contribute more
significantly to the two-body photodisintegration cross section
$\frac{d\sigma}{d\Omega}(\gamma pp \to pp)$ . At low energy, the
one-body amplitude is driven by dipole photon absorption, which is
suppressed. At high energy, it involves all other multipoles and, as a
result, the corresponding cross section remains almost constant. This
process probes the relative $pp$ wave function at a momentum which
increases with the incoming photon energy --- typically $400$ MeV/c at
$E_{\gamma}=400$ MeV, increasing to $1.5$ GeV/c at $E_{\gamma}=1.2$
GeV. Above $\sim0.8$ GeV the $pp$ wave function is not under control,
and we are reaching the limits of the model, as in the $\gamma d \to
pn$ reaction \cite{brodski}.  We may have entered a region where
quarks become the relevant degrees of freedom \cite{brodski,schulte},
or perhaps a description in terms of Regge-type calculations
\cite{grishina} is more suitable.

\subsubsection{\label{sec:res_star}Star configuration}

\begin{figure}
\includegraphics[scale=0.35]{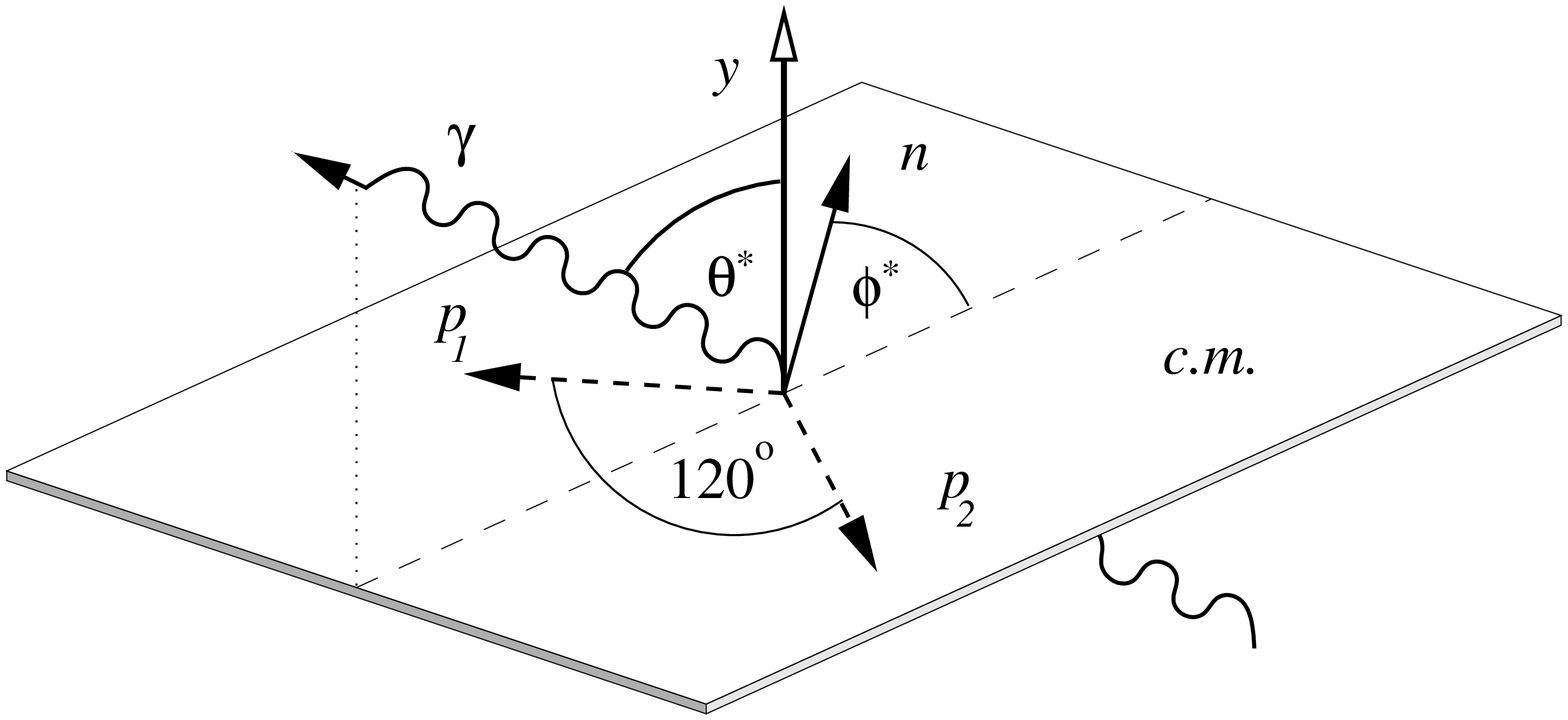}
\caption{\label{fig:kine_star} Kinematics of the star configuration in the $ppn$ center-of-mass frame. The angles $\theta^*$, between the normal vector to the star plane and the photon-beam direction, and $\phi^*$, the neutron azimuthal angle in the star plane, define the reaction.}
\end{figure}

The center of the Dalitz triangle corresponds to the three particles having equal kinetic energies and their three-momentum vectors forming angles of $120^\circ$ with each other (in the $ppn$ center-of-mass frame). For this reason, this kinematical arrangement, shown schematically in Fig.~\ref{fig:kine_star}, has been called the {\it star configuration}.
In this region, the three-body mechanisms are expected to be dominant because if the momentum is equally shared between the three nucleons, the contribution from two-body mechanisms is minimized. 
This is therefore considered to be a good place to study three-body mechanisms. 

\begin{figure}
\includegraphics[scale=0.35]{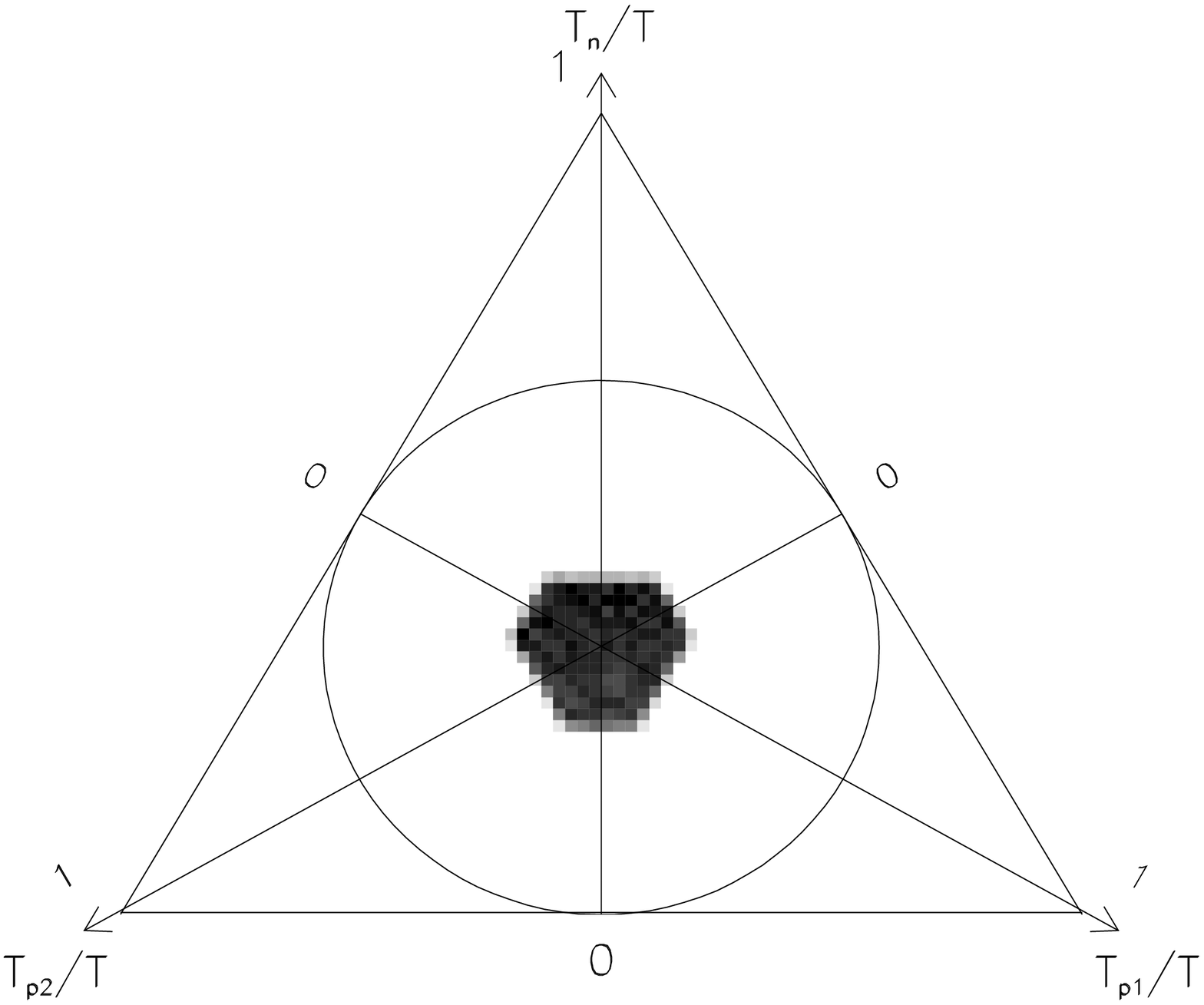}
\caption{\label{fig:dalitz_star} Dalitz plot for the CLAS $ppn$ events selected for the {\it star configuration}.}
\end{figure}

The events for this kinematics have been selected by requiring that the three nucleons satisfy the condition
\begin{eqnarray}\label{eq:star_definition}
|\theta_{ij}-120^\circ|<\delta_{\theta}
\end{eqnarray}
where
\begin{eqnarray}\label{eq:star_definition2}
\theta_{ij}={\rm arccos}\left(\frac{\overrightarrow{p_i}\cdot\overrightarrow{p_j}}{p_ip_j}\right)
\end{eqnarray}
is the angle between the momenta of nucleons $i$ and $j$, in the center-of-mass frame, and the angle $\delta_{\theta}$, which expresses the allowed deviation from the pure ``star'' kinematics, has been chosen to be $15^\circ$, as shown in Fig.~\ref{fig:dalitz_star}. 

\begin{figure}
\includegraphics[scale=0.48]{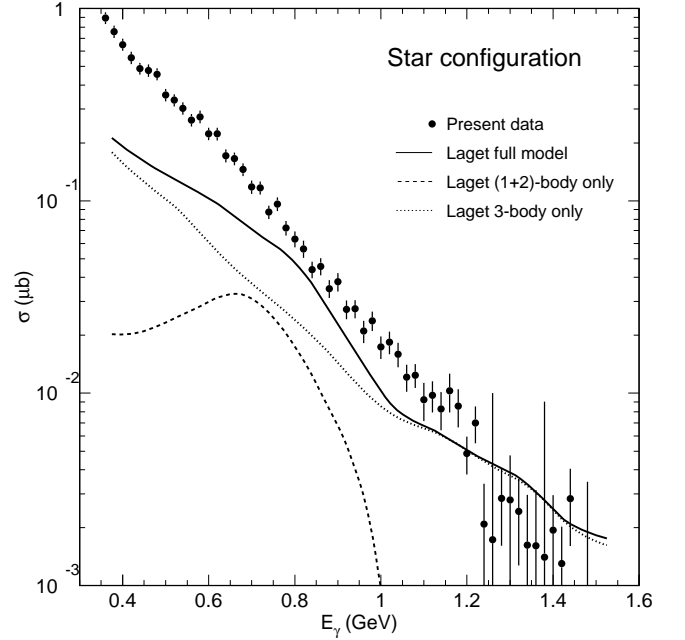}
\caption{\label{fig:star_egamma} Cross section integrated over the CLAS acceptance for the star configuration plotted as a function of incident photon energy. The CLAS data are compared with the predictions of the full model (solid curve), to the one- plus two-body-only part (dashed curve), and to the three-body-only part of the model (dotted curve).}
\end{figure}

In Fig.~\ref{fig:star_egamma}, the cross section integrated over the CLAS for the star configuration is plotted as a function of photon energy. It decreases exponentially, with slope $b\simeq 5.8$ GeV$^{-1}$, as the photon energy increases, much more steeply than for the neutron-spectator kinematics.

As expected from the kinematics, for the star configuration the
contribution of two-body mechanisms is negligible, while the bulk of
the cross section comes from three-body mechanisms. At low energy the
model misses the experimental cross section by approximately a factor
of four. The probable reason for this discrepancy is that only the
Born term and the $\Delta$-formation term \cite{bl77} have been
retained in the calculation of the $\gamma N \to \pi N$ vertex (the
upper blob in Fig.~\ref{fig:diagrams} (g) and (h)). The addition of
the contributions of the 
$N(1520)D_{13}$, $N(1440)P_{11}$, and $N(1535)S_{11}$ 
resonances also might improve the agreement with the
data.   
At high $E_{\gamma}$, the Blomqvist-Laget Born term matches the Regge
amplitudes \cite{glv97} that reproduce the $\gamma N \to \pi N$ cross
section in this energy region. The pion-rescattering amplitude
(Fig.~\ref{fig:diagrams} (g)-(i)) is parametrized in terms of partial
waves up to and including G-waves.

\begin{figure*}
\includegraphics[scale=0.5]{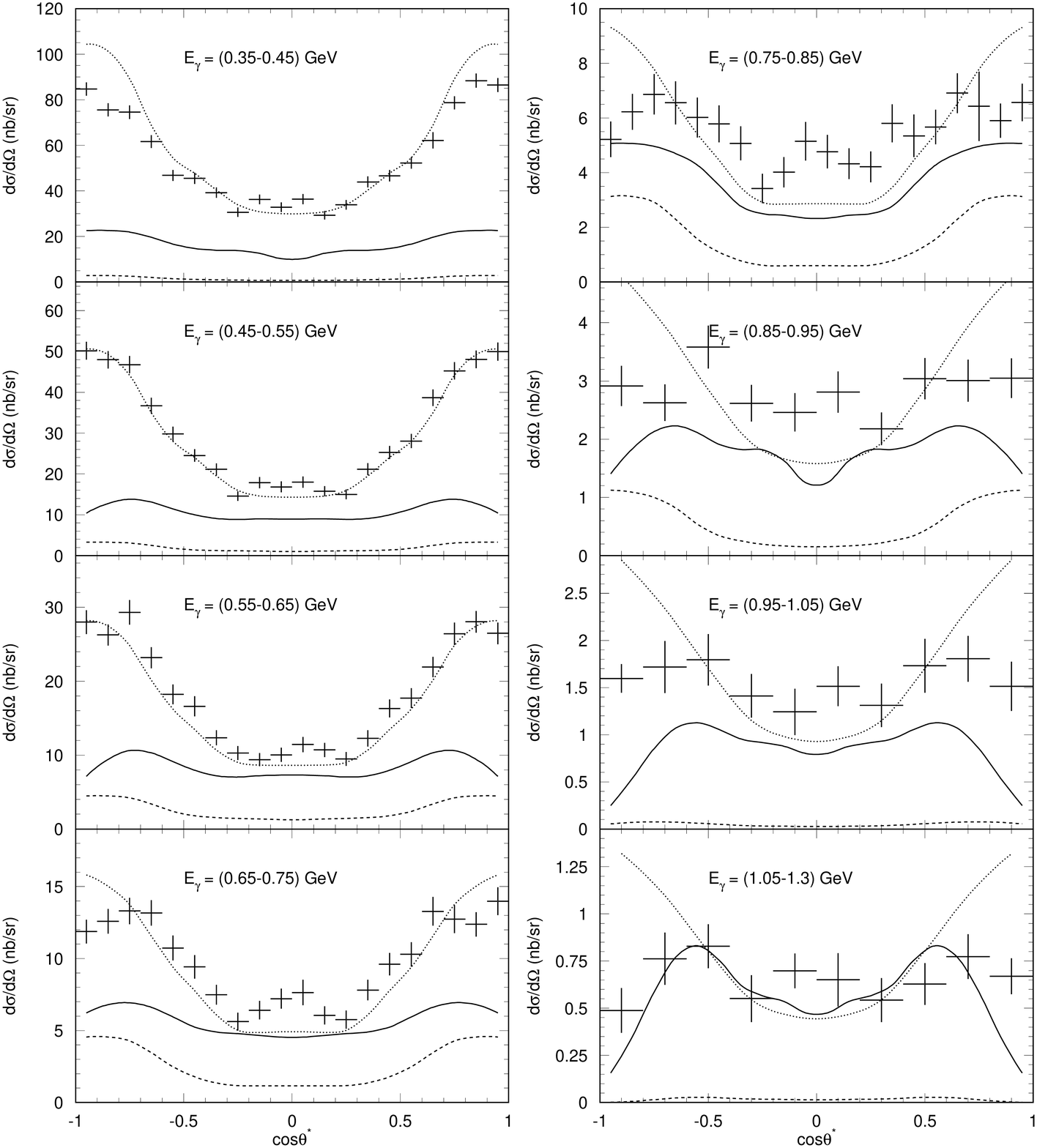}
\caption{\label{fig:star_th_1} CLAS-integrated differential cross
sections with respect to $\cos\theta^*$ for the star configuration. The
data, for photon energies between 0.35 GeV and 1.30 GeV, are compared
with the full-model results (solid curves) and the one- plus
two-body-only part (dashed curves). The dotted curves are the
phase-space distributions multiplied, for each photon-energy bin, by
the constants used to normalize the full-Dalitz cross sections.}
\end{figure*}

The differential cross section as a function of $\cos\theta^*$, the cosine of the angle between the incident photon and the normal vector to the three-nucleon center-of-mass plane (see Fig.~\ref{fig:kine_star}), is plotted in Fig.~\ref{fig:star_th_1} for eight photon-energy bins between 0.35 GeV and 1.30 GeV. 
Since the two outgoing protons are indistinguishable, the orientation of the normal vector to the star plane, $\vec p_1 \times \vec p_2$, is arbitrary. 
Thus, the distribution is symmetric around $\cos\theta^*=0$.
The shape of the cross section is very well reproduced by phase space at low energy, while at high energy the model better reproduces the curvature of the experimental distribution. At all energies the three-body mechanisms are dominant.
 
\begin{figure*}
\includegraphics[scale=0.5]{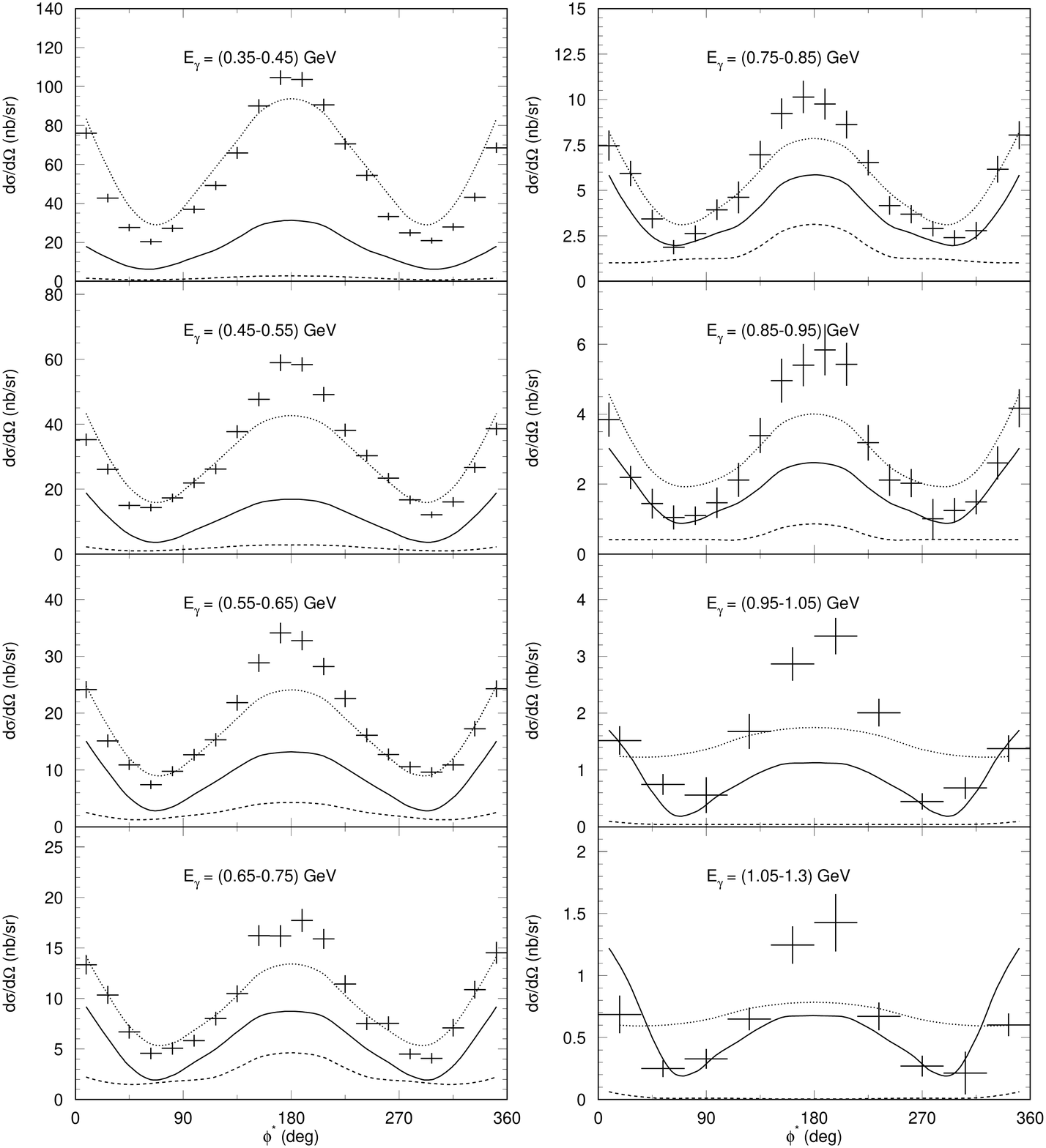}
\caption{\label{fig:star_phi} CLAS-integrated differential cross
sections with respect to $\phi^*$ for the star configuration. The data,
for photon energies between 0.35 GeV and 1.30 GeV, are compared with
the full-model results (solid curves) and the one- plus two-body-only
part (dashed curves). The dotted curves are the phase-space
distributions, multiplied, for each photon-energy bin, by the
constants used to normalize the full-Dalitz cross sections.}
\end{figure*}

Figure~\ref{fig:star_phi} shows, for eight photon-energy bins between
0.35 GeV and 1.30 GeV, the differential cross section as a function of
the angle $\phi^*$ between the neutron direction in the star plane and
the projection of the photon-beam direction in the same plane (see
Fig.~\ref{fig:kine_star}).  As is the case for $\theta^*$, the angular
distribution is symmetric, here around $180^\circ$. It also follows a
phase-space distribution, except for $E_\gamma > 0.95$~GeV,
and its shape (but
not its magnitude) is reproduced fairly well by the model as well. Again,
three-body mechanisms are seen to be dominant.

\begin{figure}
\includegraphics[scale=0.45]{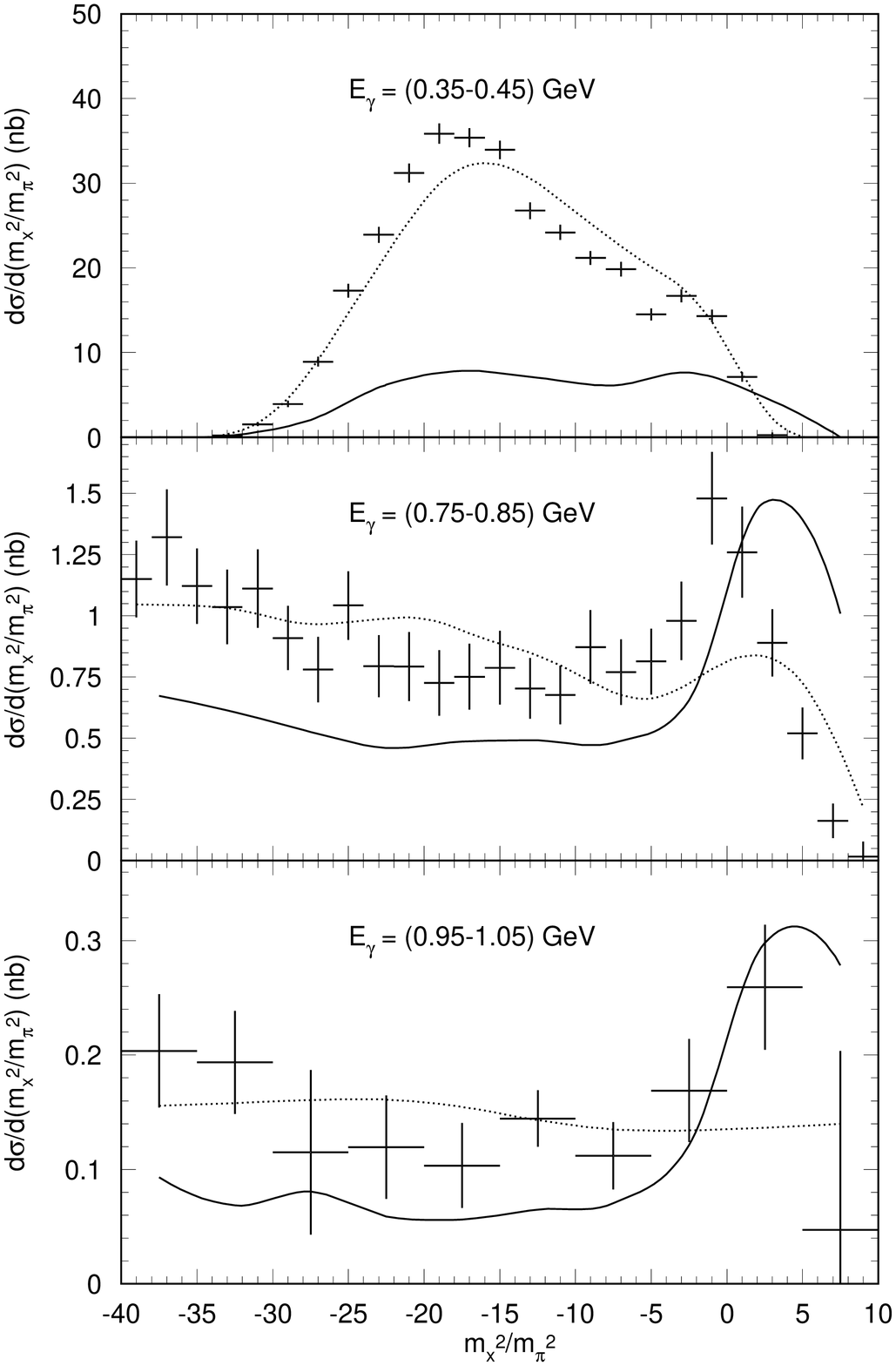}
\caption{\label{fig:star_mx}Distributions of $m_X^2/m_\pi^2$ (see Eq.~(\ref{eq:def_mx})) for the star configuration exemplified by three 0.1-GeV-wide photon-energy bins. The dotted lines represent the phase-space predictions, multiplied by the constants used to normalize the full-Dalitz cross sections, while the solid curves are the full-model results.}
\end{figure}

The photoproduced pion described by the diagrams (g) and (h) of
Fig.~\ref{fig:diagrams} can propagate on-shell, since the available
energy is larger than the sum of the masses of the pion and the three
nucleons. This causes the development of a logarithmic singularity in
the three-nucleon amplitude, which should {\it enhance} the
contribution of three-body mechanisms. The effect of this singularity
can be seen in Fig.~\ref{fig:star_mx}, in which is plotted the cross
section differential in $m_X^2/m_\pi^2$, where $m_X$, defined from the
relation
\begin{eqnarray}\label{eq:def_mx}
m_X^2 = (E_{\gamma}+m_p-E_n)^2-(\overrightarrow{k_{\gamma}}-\overrightarrow{p_n})^2,
\end{eqnarray}
is the missing mass in the $\gamma p \to \pi^+ n$ reaction, assuming that the proton is at rest.

At photon energies above about 0.6~GeV, the pion singularity appears
clearly ($m_X^2/m_\pi^2\simeq1$) in both the experimental
distributions and the model results. 
At high energy, the magnitudes of the two peaks are comparable but the shift 
of the theoretical one with respect to the experimental one reflects the 
approximate treatment of Fermi motion effects in the model. 
At lower energy, the theoretical peak is smaller than in the experiment.
The inclusion of higher-lying resonances in the sequential scattering amplitude in the model will enhance the peak near $m_X^2/m_{\pi}^2\simeq1$, but will
probably not fill the gap around $m_X^2/m_{\pi}^2\simeq-15$ for
$E_{\gamma}=400$ MeV.

These findings indicate a deviation from the sequential rescattering three-body mechanisms, which may be a hint in the search for more genuine three-body processes.

\subsubsection{\label{sec:side}Quasi-two-body breakup}

\begin{figure}
\includegraphics[scale=0.35]{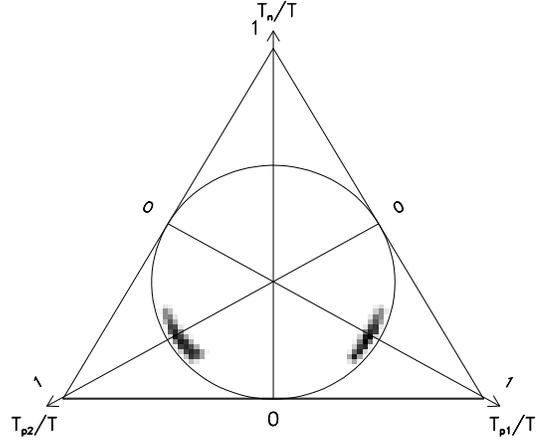}
\caption{\label{fig:dalitz_side} Dalitz plot for our CLAS $ppn$ events selected as quasi-two-body breakup.}
\end{figure}

The third region of the Dalitz plot examined corresponds to the quasi-two-body breakup, where a proton and an unbound deuteron (a $pn$ pair) are emitted back-to-back in the center-of-mass frame. For this kind of event, one of the two protons ($p_1$) is emitted with 2/3 of the total available energy, and the $pn$ pair travels in the opposite direction, with $1/3$ of the total energy, and with $T_{p2}=T_n=\frac{1}{6}T$.
This kinematics corresponds to the events in the two populated areas shown in Fig.~\ref{fig:dalitz_side}. 
These areas have been selected by requiring that the angle between the high-energy proton and each of the other two nucleons be close to $180^\circ$, and that the difference between the energies of the two low-energy nucleons be small. Using the formalism defined above,
\begin{eqnarray}\label{eq:side_definition}
|\theta_{p1p2} - 180^\circ| & < & 20^\circ, \nonumber\\
|\theta_{p1n} - 180^\circ| & < & 20^\circ, \\
{\rm and}\quad\frac{|T_{p2}-T_{n}|}{T}& < & 0.15 \nonumber
\end{eqnarray}
for the events on the right side of the Dalitz plot (where the proton labeled $p1$ has higher energy), and
\begin{eqnarray}\label{eq:side_definition2}
|\theta_{p1p2} - 180^\circ| & < & 20^\circ, \nonumber\\
|\theta_{p2n} - 180^\circ| & < & 20^\circ, \\
{\rm and}\quad\frac{|T_{p1}-T_{n}|}{T} & < & 0.15 \nonumber
\end{eqnarray}
for the events on the left side of the Dalitz plot. 
Since protons ``1'' and ``2'' are indistinguishable, the two regions of the Dalitz plot are equivalent.

\begin{figure}
\includegraphics[scale=0.48]{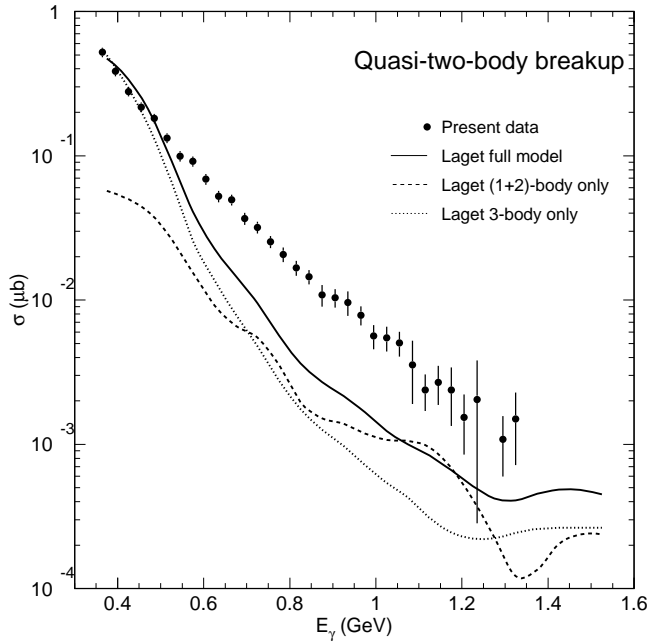}
\caption{\label{fig:side_egamma} Cross section integrated over the CLAS for the quasi-two-body breakup plotted as a function of photon energy. The data are compared with the predictions of the full model (solid curve), the (1+2)-body calculation (dashed curve), and the three-body-only calculation (dotted curve). The full-model calculation agrees quantitatively with our experimental results only up to about 0.55 GeV.}
\end{figure}

In Fig.~\ref{fig:side_egamma}, the CLAS-integrated cross section for this process is plotted as a function of photon energy. 
It decreases exponentially with a much steeper slope than for the other kinematics ($b\simeq7.3$ GeV$^{-1}$).
The full-model result is in good agreement with the experimental cross section only for the low part of the photon energy range, and seriously underestimates it above about $E_{\gamma}=0.55$ GeV. The (1+2)-body calculation gives a cross section that is smaller than the data by a factor of five or more for all photon energies.
However, this kinematic region is expected to be strongly influenced by final-state interactions (FSI) \cite{sarty92}. Only S-wave $NN$ scattering has been included in the model calculation. Furthermore, a factorization approximation has been made to estimate the nine-fold integral in Fig.~\ref{fig:diagrams}, graph (i). A full treatment, in the terms of Ref.~\cite{laget2003}, might help to reduce the discrepancy between the data and the model predictions.

It also turns out that the logarithmic singularity in the two-step
sequential scattering (Fig.~\ref{fig:diagrams} (g) and (h)) moves in
the Dalitz plot as the photon energy varies. At lower photon energies,
around $E_{\gamma}=500$ MeV, it coincides with the part of the Dalitz
plot where the quasi-two-body events are located and where the
amplitude includes a significant contribution from FSI as well. As the
photon energy increases, the singularity moves towards the top of the
Dalitz plot, and the contribution of sequential scattering to the
quasi-two-body cross section becomes negligible. Here, the difference
between the experimental cross section and the full-model result is a strong
hint of a possible contribution of other three-body mechanisms that do
not reduce to sequential scattering.

\begin{figure*}
\includegraphics[scale=0.5]{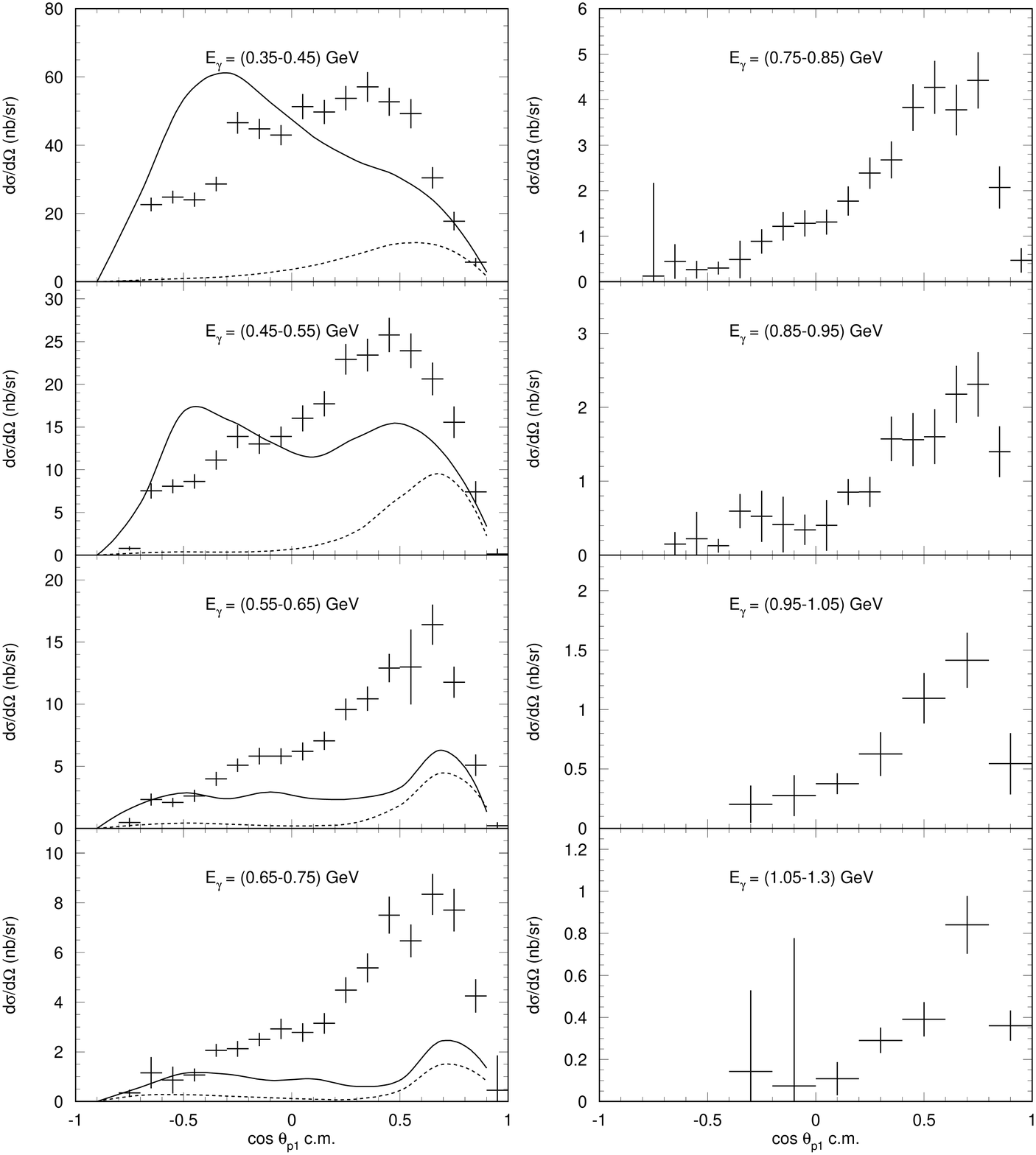}
\caption{\label{fig:side_th_p} Differential cross sections integrated
over the CLAS for the quasi-two-body breakup with respect to
$\cos\theta$ of the high-energy proton in the center-of-mass frame for photon energies between 0.35 and 1.30 GeV. Our data, for
$0.35<E_{\gamma}<0.75$ GeV, are compared with the results of the full
model (solid curves) and of the (1+2)-body-only model (dashed
curves).}
\end{figure*}

In Fig.~\ref{fig:side_th_p}, the differential cross section is plotted
as a function of the cosine of the polar angle of the 
higher-energy proton in the three-body center-of-mass frame. Data from eight
photon-energy bins between 0.35 and 1.30 GeV are shown. The
experimental cross section shows a forward peak whose relative
strength grows with increasing photon energy.  This feature is
also seen in the (1+2)-body model and in the full calculation for
$E_\gamma > 0.55$~GeV. The predicted strength of the forward peak is,
however, much too small to match the data. For lower energies, the
full calculation predicts a cross-section enhancement at backward
angles that is not seen in the data.

\subsection{The {\boldmath $ppn$} ``Three-Body" Cross Section}

Previous experiments measuring the $\gamma ^3{\mbox{He}} \rightarrow ppn$ channel in an extended part of the phase space have been performed with the DAPHNE \cite{audit97} and TAGX \cite{maruyama95} detectors.
Except for differences in the $\phi$ coverage, the CLAS event-selection cuts are very similar to the other two experiments, as seen in Table~\ref{table:cuts}; however, differences in the selection criteria of the three-body events exist between the TAGX experiment on the one hand and the DAPHNE and CLAS experiments on the other.

\begin{table*}[ht]
\renewcommand{\tabcolsep}{1pc} 
\caption{\label{table:cuts} Selection cuts applied to the TAGX, DAPHNE, and CLAS $\gamma ^3\mbox{He} \to ppn$ experiments in order to extract the ``three-body" total cross section.}
\begin{center}
\begin{tabular}{ c  c  c }
\hline\hline
{TAGX} & {DAPHNE} & {CLAS}\\
\hline
{$15^{\circ} \leq \theta_{p_1,p_2} \leq 165^{\circ}$} & {$22^{\circ} \leq \theta_{p_1,p_2} \leq 158^{\circ}$} & {$15^{\circ} \leq \theta_{p_1,p_2} \leq 125^{\circ}$}\\
{$0^{\circ} \leq \phi_{p_1,p_2} \leq 40^{\circ}$} & {$0^{\circ} \leq \phi_{p_1,p_2} \leq 360^{\circ}$} & {CLAS $\phi$ fiducial cuts}\\
{$p_{p_1,p_2}\geq300$ MeV/c} & {$p_{p_1,p_2}\geq300$ MeV/c} & {$p_{p_1,p_2}\geq300$ MeV/c}\\
{``Non-spectator'' neutron} & {$p_n\geq 150$ MeV/c} & {$p_n\geq 150$ MeV/c}\\
\hline\hline
\end{tabular}
\end{center}
\end{table*}

The $ppn$ ``three-body'' cross section is defined as:
\begin{eqnarray}\label{eq:sigma_tot_abs}
\sigma_{3body}(E_{\gamma})=\frac{N_{3body}(E_{\gamma})}{N_{\gamma}(E_{\gamma})\,Acc(E_{\gamma})\,\frac {\rho z N_A}{A}},
\end{eqnarray}
where $N_{3body}$ is the number of events extracted by applying the
selection cuts given in Table~\ref{table:cuts} and $Acc$ is the
acceptance of the CLAS detector for the $ppn$ events calculated with
the phase-space Monte-Carlo simulation.  The low-momentum neutrons
($p_n \le 150$ MeV/c) have been excluded in order to select only those
events for which all three particles participate in the reaction, thus
diminishing the importance of two-body processes
\cite{maruyama95,audit97}. In this kinematics, the phase-space result
describes the process reasonably well.

\begin{figure}
\includegraphics[scale=0.45]{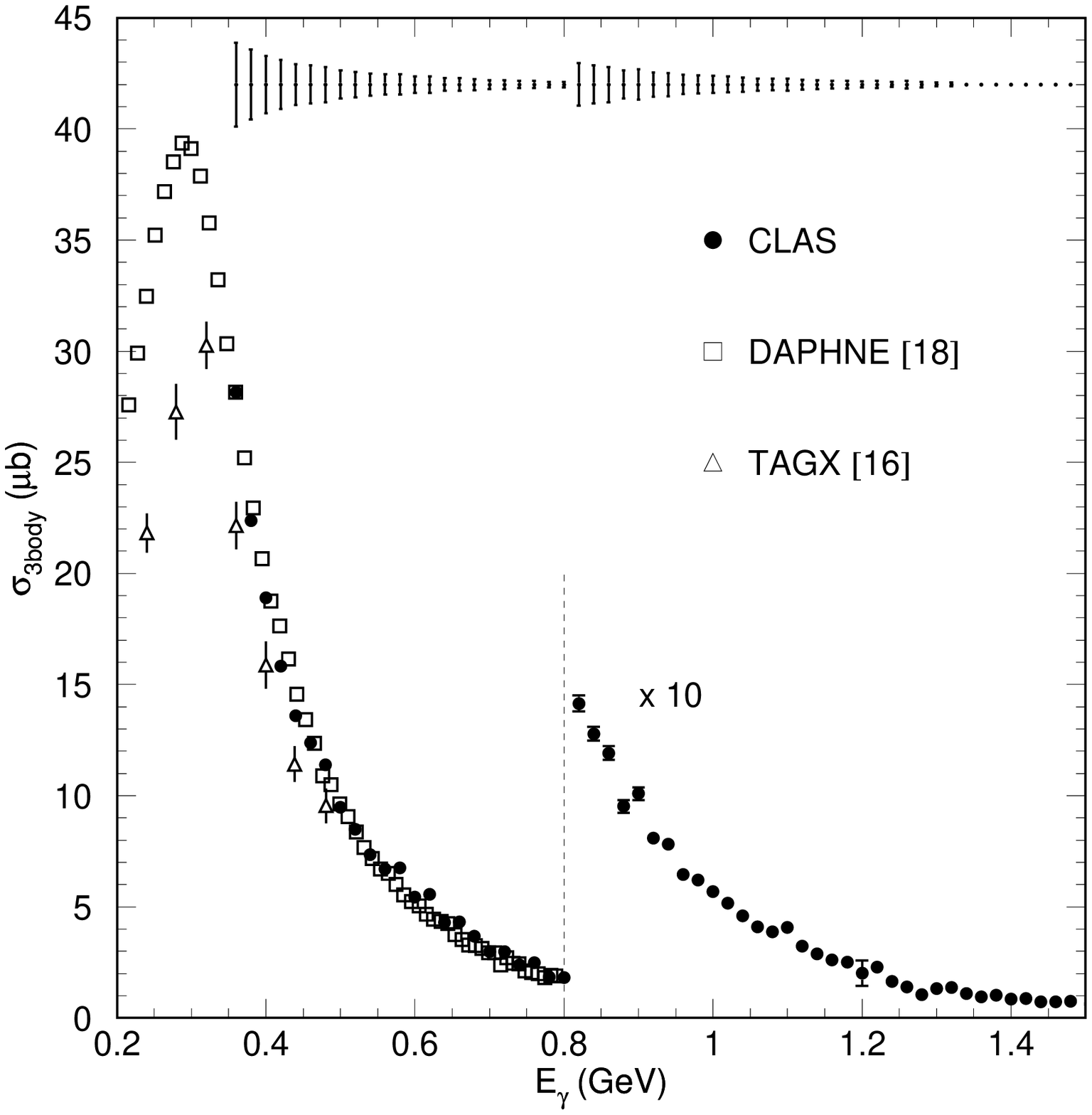}
\caption{\label{fig:daphne} Total ``three-body" cross section as defined by Eq.~(\ref{eq:sigma_tot_abs}) for the $\gamma ^3{\mbox{He}} \to ppn$ reaction plotted as a function of photon energy. The CLAS data (full circles) are compared with the results from DAPHNE \cite{audit97} (empty squares) and TAGX \cite{maruyama95} (empty triangles). The error bars on the CLAS experimental points are statistical only. The CLAS systematic uncertainties are represented by the vertical bars in the upper part of the figure.}
\end{figure}

Figure~\ref{fig:daphne} shows $\sigma_{3body}$ as a function of the
photon energy $E_{\gamma}$.  The full circles represent our CLAS data,
the empty triangles the data of the TAGX Collaboration
\cite{maruyama95}, and the empty squares the results obtained in the
experiment carried out at MAMI with the DAPHNE detector
\cite{audit97}. The error bars on the CLAS data are statistical
only. The systematic uncertainties delineated in the previous section
are shown by the vertical lines in the upper part of the figure.

In the overlap region of the three experiments from 0.35 to 0.80 GeV, the CLAS data are in good agreement with the DAPHNE results, but differ from the TAGX cross sections by about 15\%, most likely due to the above-mentioned difference in the three-body event selection. Above 0.80 GeV, no previous data are available.

The phase-space extrapolation to the unmeasured regions has been done only for comparison with the previous experiments, which adopted the same procedure to extract $\sigma_{3body}$. 

\section{\label{summary}SUMMARY AND CONCLUSIONS}

The three-body photodisintegration of $^3$He has been measured with
the tagged-photon beam and the CEBAF Large Acceptance Spectrometer in
Hall B at the Thomas Jefferson National Accelerator Facility in the
photon-energy range between 0.35 and 1.55 GeV. This measurement
constitutes a wide-ranging survey of two- and three-body processes in
the $\gamma ^3{\rm He} \rightarrow ppn$ reaction channel, as a
consequence of the high statistics and large kinematic coverage
obtained with the CLAS.

Total and partially integrated differential cross sections for the
full $ppn$ data set and for selected kinematics were extracted and are
compared with phase-space distributions and with the predictions of
the diagrammatic model of Laget. This model reproduces some of the
main trends of the experimental energy distributions, and for these
cases can be taken as a qualitative guide to understanding the
reaction mechanisms.

From the analysis of the neutron-momentum distribution for the full
Dalitz plot, the kinematic region corresponding to the
photodisintegration of a $pp$ pair in the presence of a spectator
neutron has been identified. Here, the effects of two-body absorption
mechanisms dominate and the model results are 
very close to experiment at low energy, up to $E_{\gamma}$~=
600~MeV. At higher energies, the discrepancy, which increases
with energy, might be a hint that we are approaching the limit
of models based on meson and baryon degrees of freedom.

A strong contribution of three-body sequential meson-absorption
mechanisms is manifested over all the available phase space, but most
especially in the {\it star} kinematics, the spatially symmetric
configuration of the three final-state nucleons. These events
are dominated by the coupling to the $\Delta$ resonance, and they
strongly confirm its role in three-body forces. The deviations from the
predictions of the diagrammatic model point not only toward the
necessity of implementing processes which involve higher-lying
baryonic resonances, but also toward possible additional three-body
mechanisms beyond sequential scattering.

The $4\pi$-integrated ``three-body" cross section is in excellent agreement with previous experimental results from DAPHNE up to 800 MeV. For the first time we now have provided access to a higher energy range, up to 1.5 GeV.

This work breaks new ground in the experimental study of the three-body photodisintegration of $^3$He. However, before making contact with the elusive three-body forces, it calls for a more complete treatment of three-body mechanisms which go beyond the dominant sequential meson exchange and $\Delta$ formation in the intermediate energy range, and which take into account possible coupling with partonic degrees of freedom in the highest energy range.

\begin{acknowledgments}
We would like to thank the staff of the Accelerator and Physics
Divisions at Jefferson Lab, who made this experiment possible. Acknowledgments
for the support of this experiment go also to the Italian Istituto
Nazionale di Fisica Nucleare, the French Centre National de la
Recherche Scientifique and Commissariat a l'Energie Atomique, the
U.S. Department of Energy and the National Science Foundation, and the
Korea Science and Engineering Foundation. Southeastern Universities
Research Association (SURA) operates the Thomas Jefferson National
Accelerator Facility under U.S. Department of Energy contract
DE-AC05-84ER40150. The GWU Experimental Nuclear Physics Group is 
supported by the U.S. Department of Energy under grant DE-FG02-95ER40901.
\end{acknowledgments}

\bibliography{paper_v05}
\end{document}